\documentclass[%
 reprint,
superscriptaddress,
nofootinbib,
 amsmath,amssymb,
 aps,
 pra,
 twocolumn,
]{revtex4}

\usepackage{graphicx}
\usepackage{bm}
\usepackage[colorlinks, linkcolor=blue, citecolor=blue, urlcolor=blue, breaklinks=false]{hyperref}
\graphicspath{{figures/}} 
\usepackage{color}

\newcommand{\mi}{{\rm i}}
\newcommand{\me}{{\rm e}}
\newcommand{\id}{\mathbb{1}}
\newcommand{\llangle}{\langle\!\langle}
\newcommand{\rrangle}{\rangle\!\rangle}
\usepackage{braket}
\usepackage{amsmath}
\usepackage{amsfonts}
\usepackage{amssymb}
\usepackage{bbold}
\usepackage{float}
\usepackage{float}

\allowdisplaybreaks
\begin{document}

\title{A generalized phase space  approach for solving quantum spin dynamics}

\author{Bihui Zhu}
\address{ITAMP, Harvard-Smithsonian Center for Astrophysics, Cambridge, MA 02138, USA}
\address{Department of Physics, Harvard University, Cambridge, MA 02138, USA}

\author{Ana Maria Rey}
\address{JILA, NIST, and Department of Physics, University of Colorado, 440 UCB, Boulder, CO 80309, USA}

\address{Center for Theory of Quantum Matter, University of Colorado, Boulder, CO, 80309, USA}

\author{Johannes Schachenmayer}
\address{CNRS, IPCMS (UMR 7504), ISIS (UMR 7006), and Universit\'{e} de Strasbourg, 67000 Strasbourg, France}

\begin{abstract}
Numerical techniques to efficiently model out-of-equilibrium dynamics in interacting quantum many-body systems are key for advancing our capability to harness and understand complex quantum matter. Here we propose a new numerical approach which we refer to as GDTWA. It is based on a discrete semi-classical phase space sampling and allows to investigate quantum dynamics in lattice spin systems with arbitrary $S\geq 1/2$. We show that the GDTWA can accurately simulate dynamics of large ensembles in arbitrary dimensions. We apply it for $S>1/2$ spin-models with dipolar long-range interactions, a scenario arising in recent experiments with magnetic atoms. We show that the method can capture beyond mean-field effects, not only at short times, but it also can correctly reproduce long time quantum-thermalization dynamics. We benchmark the method with exact diagonalization in small systems, with perturbation theory for short times, and with analytical predictions made for models which feature quantum-thermalization at long times. We apply our method to study dynamics in large $S>1/2$ spin-models and compute experimentally accessible observables such as Zeeman level populations, contrast of spin coherence, spin squeezing, and entanglement quantified by single-spin Renyi entropies. We reveal that large $S$ systems can feature larger entanglement than corresponding $S=1/2$ systems. Our analyses demonstrate that the GDTWA can be a powerful tool for modeling complex spin dynamics in regimes where other state-of-the art numerical methods fail.
\end{abstract}

\maketitle

\section{\label{sec:intro}Introduction}

In the past years, rapid developments of various experimental platforms have made it possible to observe out-of-equilibrium dynamics of large isolated quantum many-body models in controlled environments \cite{Bloch_Quantum_2018, Cirac_Goalsa_2012, Bloch_Quantum_2012, Blatt2012}. Naturally, this also leads to a high demand for numerical methods capable of simulating such dynamics. Computations for large system sizes beyond a classical mean-field picture are a challenging task due to the complexity of the full quantum problem. Consequently, most recently developed methods for large systems are limited to either low dimensional geometries (e.g.~by making use of matrix product states/tensor networks~\cite{Orus_Apract_2014, Schollwock_Theden_2011, Verstraete_Matrix_2008, Vidal_Efficie_2003, White_Real-Ti_2004, Daley_Time-de_2004}), particular ansatz wave-functions (e.g.~in combination with variational Monte-Carlo evolution~\cite{Carleo_Localiz_2012,Cevolani_Protect_2015, Ido_Time-de_2015, Carleo_Solving_2017}), or dilute systems (e.g.~by making use of a clusterization~\cite{Hazzard2014b,Orioli2017}). 

In particular, models of coupled spin-particles with long-range interactions have become a topic of intensive research because of important experimental progress. While models with spin $S=1/2$ have been implemented with many different setups, e.g.~using polar molecules~\cite{Yan2013,Hazzard2014b}, Rydberg atoms~\cite{labuhn2016,Zeiher2017,Bernien2017,Barredo2018,Guardado2018,Takei_Direct_2016, Orioli2017}, trapped ions~\cite{Garttner2017,Bohnet2016, Neyenhuise2017} and cavity QED systems~\cite{Norcia2018,Davis2019}, recently also models with larger spins $S>1/2$ have become a research focus in particular for experiments with magnetic atoms~\cite{Lepoutre_Out-of-_2019,Baier2016,dePazMISF,depaz2013,depaz2016,manfred2019}.  The large spin degrees of freedom in  $S>1/2$ systems poses a much more stringent requirement for numerical treatment compared to $S=1/2$ systems. The full Hilbert space involves $(2S+1)^N$ states for $N$ particles, which renders exact diagonalization (ED) impossible already for small $N$. In addition, many current experiments are performed in 2D or 3D, thus also disabling otherwise very successful matrix product state techniques for long-range interacting systems in 1D~\cite{Zaletel_Time-ev_2015,Haegeman_Unify_2016,Schachenmayer_Entangl_2013,Hauke_Spread_2013}. This calls for new theoretical tools capable of accounting for the many-body nature of these models as well as intrinsic quantum correlations. 

In this paper, we present an efficient numerical approach based on a semi-classical phase space method that can be applied to large 2D/3D lattice systems with arbitrary spin $S$. The method is based on the well known truncated Wigner approximation (TWA) \cite{Steel_Dynamic_1998,Blakie_Dynamic_2008,Polkovnikov} adapted to spin-models. The general TWA idea relies on a sampling of the quantum fluctuations of the initial state from a Wigner function, and an evolution of the samples along classical trajectories. Importantly, in contrast to previous approaches, here we introduce an enlarged phase space representing not only $3$ spin-variables, but all $(2S+1)^2$ density matrix elements for each spin \cite{Lepoutre_Out-of-_2019, anatolicluster2018,manfred2019,Davidson2015}. Furthermore,  we can use a sampling from discrete quasi-probability distributions, which are exact and positive for most experimentally relevant initial states. Our method thus allows us to study the TWA time-evolution not only of spin operators, but the full spin-density matrix and thus allows us to extract experimentally relevant time-dependent observables such as spin-state populations, and fundamentally relevant quantities such as entanglement. In the limit of $S=1/2$ our generalized discrete TWA approach (GDTWA), reduces to the previously proposed discrete TWA method (DTWA) \cite{Schachenmayer2015a}, which has been remarkably succesful in predicting $S=1/2$ model dynamics \cite{Schachenmayer2015b,Pucci_Simul_2016,PineiroOrioli_Nonequi_2017,Acevedo_Explori_2017,Orioli2017,Czischek_Quenche_2018}.

This paper is organized as follows: In Sec.~\ref{sec:method}, after a brief review of the TWA for continuous Wigner functions, we introduce the GDTWA. In Sec.~\ref{sec:benchmark}, we test its validity by a comparison with ED. We focus on the evolution of Zeeman level populations induced by dipolar interactions  (relevant to experiments using both Cr \cite{Lepoutre_Out-of-_2019,dePazMISF,depaz2013,depaz2016} and Er atoms~\cite{manfred2019}) and the evolution of entanglement. In Sec.~\ref{sec:ent}, we apply the GDTWA to investigate  various aspects of spin dynamics: the spreading of population in a synthetic dimension and the underlying approach to thermalization, as well as the build-up of quantum entanglement. Finally, in Sec.~\ref{sec:outlook} we  conclude and discuss applications for other systems as an outlook.

\section{Method}\label{sec:method}
\subsection{Phase space sampling for quantum spin systems}

The Hamiltonian for a system consisting of $N$ spin-$S$ particles coupling to each other via two-body interactions can be written as
\begin{align}
   \hat H_S = \sum_{i,\alpha} u_\alpha^i \hat S_{\alpha}^i + \sum_{i\neq j,\alpha,\beta} w^{i,j}_{\alpha,\beta} \hat S_{\alpha}^i\hat S_{\beta}^j
   \label{eq:S_ham},
\end{align}
with $\alpha,\beta=x,y,z$, and $i,j=1,2,...N$. The first term governs local fields and the second term inter-spin interactions. The dynamics of the three components of the  spin-operator  of  the  $i$ particle, $\langle \hat S^i_{\alpha = x,y,z}\rangle$, can be obtained via the Heisenberg equations of motion (we set $\hbar \equiv 1$ throughout this paper):
\begin{align}
    \frac{d}{dt} \langle \hat S_\alpha^i \rangle = {\mi}\langle [\hat H_S,\hat S_\alpha^i] \rangle.
    \label{eq:heis}
\end{align}
These equations  generally  depend on high order inter-spin correlations such as $ \langle\hat S_\alpha^i \hat S^j_\beta\rangle$, $ \langle\hat S_\alpha^i \hat S^j_\beta \hat S^k_\gamma \rangle$, $\dots$, for $i\neq j \neq k$ and in practice it is impossible to solve them exactly for large $N$. In a mean-field ansatz one assumes that such spin correlations factorize and can be written in terms of single particle observables, $\langle\hat S_\alpha^i\rangle$, e.g., $ \langle\hat S_\alpha^i \hat S^j_\beta\rangle \approx \langle\hat S_\alpha^i\rangle \langle\hat S_\beta^j\rangle$. Then Eqs.~\eqref{eq:heis} turn into $N$ coupled closed equations which can be easily solved numerically, and which correspond to equations for classical spin-variables\footnote{Note that more general models than Eq.~\eqref{eq:S_ham} for $S>1/2$ can also depend on intra-spin terms such as quadratic fields $\propto (\hat S_z^i)^2$. In this case also the mean-field equations are not necessarily closed unless also intra-spin correlations are assumed to factorize. Those cases are not accounted for in the approach described here, but we will properly include them in our GDTWA method below.}.  Importantly, the factorization neglects entanglement between the spins, as the total state of the system is forced to remain a product-state. The missing quantum correlations are in many cases crucial. One general approach to account for some of those correlations is to retain higher order correlations using e.g.~BBGKY hierarchies~\cite{Pucci_Simul_2016}. However, then already retaining second-order correlations increases the complexity of Eqs.~\eqref{eq:heis} to $\sim N^2$ and furthermore higher order corrections can typically lead to numerical instabilities. An alternative approach to simulate quantum correlations while retaining a complexity $\sim N$ is to resort to a phase space description of the quantum system \cite{Blakie_Dynamic_2008,Polkovnikov} as reviewed in the following.

In the phase space approach, quantum operators $\hat O$ are mapped to functions in the phase space of classical variables, so-called Weyl symbols. Taking as an example a particle moving in 1D, the phase space variables are given by position and momentum $x$ and $p$, respectively, and the Weyl symbol is denoted as $O_W(x,p)$.  The expectation value of any operator $\hat O$ can then be computed as an integral in phase space:
\begin{align}
\langle\hat O\rangle = {\rm Tr}[\hat \rho \hat O]=\int dx\,dp\, W(x,p)O_W(x,p).\label{eq:OW}
\end{align}
Here, $W(x,p)$ is the Wigner function corresponding to the Weyl symbol of the density matrix $\hat\rho$: $W(x,p)= 1/(2\pi) \int ds\, \bra{x+s/2}\hat\rho\ket{x-s/2} \exp(-\mi ps)$, and represents a quasiprobability distribution for points in phase space. In the truncated Wigner approximation (TWA), the time evolution of $\langle\hat O\rangle$ is approximated from the initial Wigner function and the classical evolution of the phase space variables:
\begin{align}
\langle\hat O\rangle(t) \approx \int dx_0\,dp_0\, W(x_0,p_0)O_W(x_{\rm cl}(t),p_{\rm cl}(t)).
\end{align}
Here, $x_{\rm cl}(t)$ and $p_{\rm cl}(t)$ are classical trajectories of $x$ and $p$, respectively, with $x_{\rm cl}(0) = x_0$ and $p_{\rm cl}(0)=p_0$. Then, compared to a fully classical evolution, in the TWA dynamics quantum fluctuations are taken into account to the lowest order, in the sense of keeping only terms linear in $\hbar$ in the equations of motions for the Weyl symbols~\cite{Polkovnikov,Blakie_Dynamic_2008}. Observables (hermitian operators) give rise to real Weyl symbols and thus correspond to symmetrized sums of the position and momentum operator. For example, the Weyl symbol $O_W(x_{\rm cl}(t), p_{\rm cl} (t))= x_{\rm cl}^2(t) p_{\rm cl}(t)$ corresponds to the observable $\hat O  = (\hat x^2 \hat p + 2\hat x \hat p \hat x + \hat p \hat x^2)/2$ \cite{Polkovnikov}.

\begin{figure*}
\centering
\includegraphics[width=0.7\textwidth]{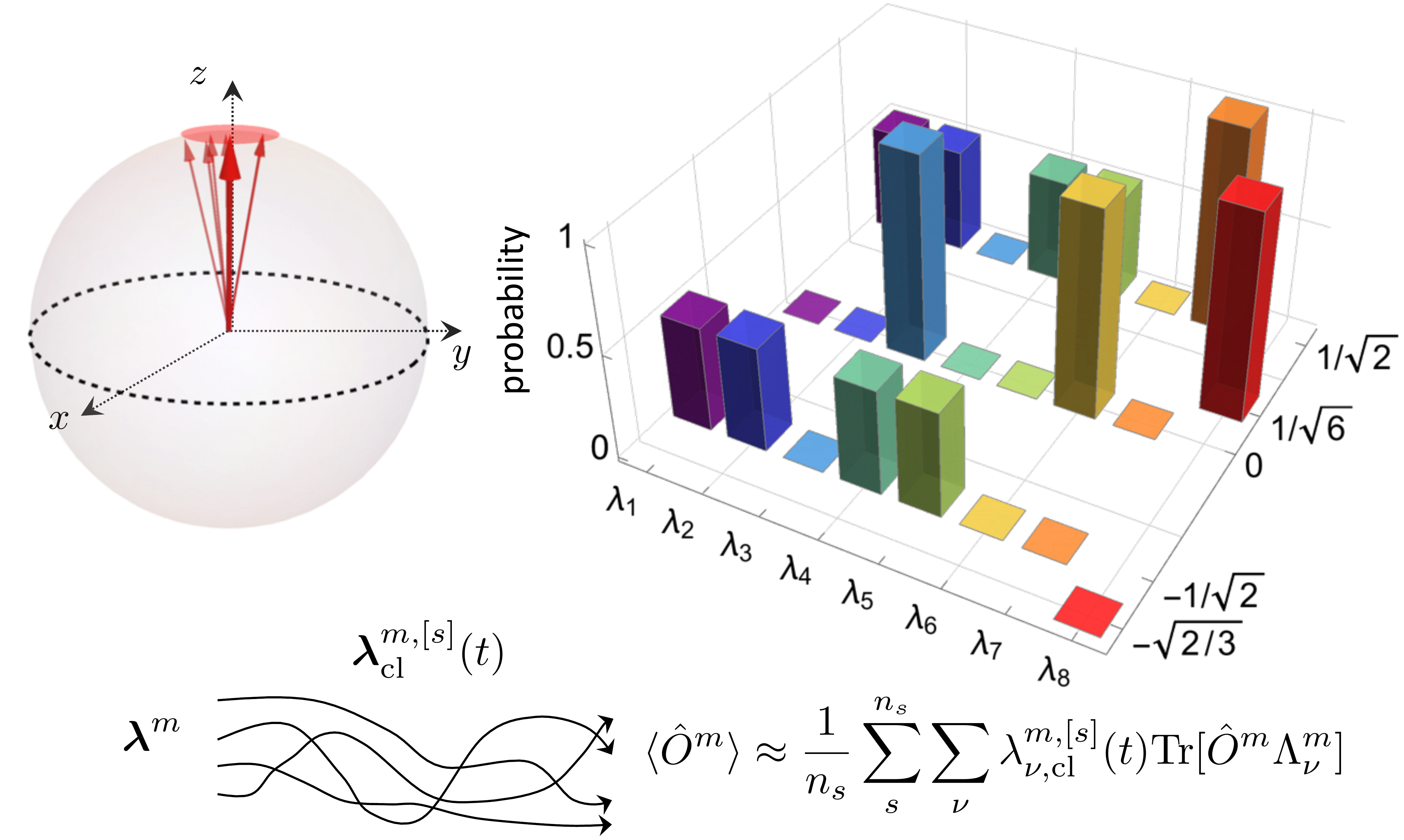}
\caption{\label{fig:scheme} In a conventional TWA approach for spin systems, an initial single spin-$S$ prepared in a spin coherent state polarized along $z$ can be represented in a continuous 3D phase space, i.e.~by a fluctuating 3D vector. Its mean direction points along $z$ and uncertainties in the orthogonal directions have a width $\sim\sqrt{S}$. In contrast, in the GDTWA approach the state of a single spin $m$ is represented by a fluctuating $(\mathcal{D}-1)$-dimensional Bloch vector, ${\bm{\lambda}}^m$, with $\mathcal{D}=(2S+1)^2$, e.g.~$\mathcal{D}-1=8$ is shown for $S=1$. States are described by discrete probability distributions, i.e.~each vector component $\lambda_\nu^m$ can only assume certain discrete values with corresponding probabilities. To simulate the time-evolution we sample the initial generalized Bloch vector elements from this discrete distribution. Observables at later times are then approximated by statistically averaging classically evolved samples (see text).}
\end{figure*}

For a system of $N$ coupled spin-$S$ particles, a natural way to formulate the TWA dynamics consists of replacing $\{\hat{x},\hat{p}\}\rightarrow \{\hat S^i_\alpha\}$, and using the $3N$ spin-variables $\{S^i_\alpha\}$ (with $\alpha =x,y,z$ and $i = 1,2,\dots N$), corresponding to the Weyl symbols of $\{\hat S^i_\alpha\}$, as phase space variables. The trajectories $S^i_{\alpha, {\rm cl}}(t)$ correspond to those obtained from the mean-field approximation of Eq.~\eqref{eq:heis} with $S_{\alpha, {\rm cl}}^i \equiv \langle \hat S^i_\alpha\rangle$. For particular initial states, the Wigner function can be easily found. Taking for example a product state $\prod_i \hat \rho^i$, where each spin $i$ is described by a spin coherent state, then for large $S$ the Wigner function can be well approximated by a Gaussian. In the case of a state initially polarized along the $z$ direction, the Wigner function factorizes, and for each spin takes the simple form
\begin{align}
W(S_x^i,S_y^i,S_z^i)=\frac{1}{\pi S}\me^{ -[(S_x^i)^2+(S_y^i)^2]/S}\delta(S_z^i-S),
\label{eq:gaussian_zstate}
\end{align}
where $\delta(\cdot)$ denotes the delta function. For this state the phase space value of $S^i_z$ is determined as $S$, while the values of $S^i_x$ ($S^i_y$) fluctuate according to a Gaussian distribution with zero mean and a variance $\Delta (S_x^i)^2=\Delta (S_y^i)^2=S/2$ (see Fig.~\ref{fig:scheme}). This width of the distribution of the classical variables reflects the uncertainty relation intrinsic to the quantum mechanical operators $\hat S^i_x$ and $\hat S^i_y$. For large $S\gg 1$,  $\Delta S^i_x/S\to 0$,  suggesting vanishing effects from quantum fluctuations. 

For such initial states, the TWA evolution now reads
\begin{align}
\langle\hat O\rangle(t) \approx \int \prod_{i,\alpha} d S_{\alpha,0}^i W(S_{\alpha,0}^i) 
O_W(\{S_{\beta, \rm{cl}}^j(t)\}),
\label{eq:TWA_gaussian}
\end{align}
where the Weyl symbol corresponds again to symmetrized observables. For example, for an inter-spin correlation such as $\hat O = \hat S_x^n \hat S_x^m$ at distinct sites $n\neq m$, simply $O_W(\{S_{\beta, \rm{cl}}^j(t)\})= S_{x,{\rm cl}}^n(t) S_{x, {\rm cl}}^m(t)$. It is important to note that, in contrast to a mean-field simulation, a TWA evolution can lead to inter-spin quantum correlations. For example, due to the time-evolution according to (generally) non-linear classical equations, a spin-component for a single spin $m$ can now depend on the initial conditions of {\em all} other spin-variables via some function, e.g.~$S_{x,{\rm cl}}^m (t)\equiv \mathcal{F}_x^m (\{S_{\beta,0}^j\})$. Therefore, for observables computed in the TWA, such as $\hat S^m_x \hat S^n_x$ with $m\neq n$,
\begin{align}
\langle\hat S^m_x \hat S^n_x\rangle(t) &\approx \int \prod_{i,\alpha} dS_{\alpha,0}^i W(S_{\alpha,0}^i) 
\mathcal{F}_x^m (\{S_{\beta,0}^j\})
\mathcal{F}_x^n (\{S_{\gamma,0}^k\}) \nonumber\\
&\neq \langle\hat S^m_x\rangle(t) \langle\hat S^n_x\rangle(t),
\end{align}
and correlations can build up with time. Generally, the factorization in the last equation would only hold if $\mathcal{F}_x^m$ would only depend on the variables of the spin $m$ itself.

\subsection{Generalized discrete Truncated Wigner approximation (GDTWA)}

There are some important shortcomings of the Gaussian TWA approach with $3$ spin-variables introduced in the previous section. Most importantly, the scheme is tailored to problems relying on the $3$ spin operators, both in the Hamiltonian and for measurements. Most experimental scenarios go beyond this limitation, by including e.g.~quadratic fields $\propto (\hat S_z^i)^2$ and measurements of Zeeman state populations~\cite{Lepoutre_Out-of-_2019,manfred2019}. Furthermore, the approach is insufficient for a full state-tomography in large $S$ systems (see~\ref{app:gaussian_tom}). More generally, we would like to adapt the method to arbitrary many-body models of coupled discrete local Hilbert spaces.

To go beyond the limitations we proceed by enlarging our phase space from $3$ spin-variables to elements of higher-dimensional generalizations of Bloch vectors \cite{Bertlmann_Blochv_2008}. This can be accomplished by noticing that for  a spin-$S$ atom with  $\mathcal{N}=2S+1$ spin states, its density matrix $\hat \rho_i$ consists of $\mathcal{D}=\mathcal{N}\times\mathcal{N}$ elements. Correspondingly, we can define $\mathcal{D}$ hermitian operators, $\Lambda_\mu$ (hats are dropped for clarity of notation),  using the generalized Gell-Mann matrices (GGM)  and the identity matrix $\id$\footnote{Note that for convenience, we use a slightly different normalization factor than in the definition of usual (generalized) Gell-Mann matrices \cite{Bertlmann_Blochv_2008}, which will simplify operator expansions.}:
 \begin{align}
\Lambda^i_{\mu=1, \dots,\mathcal{N}(\mathcal{N}-1)/2}&=\frac{1}{\sqrt{2}}(\ket{\beta}\bra{\alpha}_i+{\rm h.c.})\nonumber\\
&\quad\text{for}\quad \alpha>\beta,\quad 1\le \alpha,\beta\le \mathcal{N},\label{eq:Rs}\\
\Lambda^i_{\mu=\mathcal{N}(\mathcal{N}-1)/2+1, \dots, \mathcal{N}(\mathcal{N}-1)} &=\frac{1}{\sqrt{2}\mi}(\ket{\beta}\bra{\alpha}_i-{\rm h.c.})\nonumber\\
&\quad\text{for} \quad \alpha>\beta,\quad 1\le \alpha,\beta\le \mathcal{N},\label{eq:Is}\\
\Lambda^i_{\mu=\mathcal{N}(\mathcal{N}-1)+1, \dots, \mathcal{N}^2-1}
                                                     &=\frac{1}{\sqrt{\alpha(\alpha+1)}}\times \nonumber\\
   &\sum_{\beta=1}^\alpha \left(\ket{\beta}\bra{\beta}_i-\alpha\ket{\alpha+1}\bra{\alpha+1}_i \right)\nonumber\\
&\text{for} \quad 1\leq\alpha<\mathcal{N},\label{eq:Ds}\\
\Lambda^i_{\mathcal{D}} &= \sqrt{\frac{1}{\mathcal{N}}} \id_i.
\end{align}
For each spin $i$, these matrices are orthonormal, ${\rm Tr}[\Lambda^i_\mu\Lambda^i_\nu]=\delta_{\mu,\nu}$, and constitute a complete local basis. Hence any single-spin operator can be represented via
\begin{align}
    \hat O^i&=\sum_\mu c^i_\mu \Lambda^i_\mu, \quad \text{with} \quad
 c^i_\mu={\rm Tr}[\Lambda^i_\mu\hat O^i],\label{eq:ggm_exp}
\end{align}
and  $\mu=1,2,...,\mathcal{D}$. Consequently, more general Hamiltonians than Eq.~\eqref{eq:S_ham}, with one- and two-body terms can now be represented as 
\begin{align}
   \hat H_\Lambda = \sum_{i,\mu} u_\mu^i \Lambda_{\mu}^i + \sum_{i,j\neq i,\mu,\nu} w^{i,j}_{\mu,\nu} \Lambda_{\mu}^i\Lambda_{\nu}^j,
   \label{eq:Lambda_ham}
\end{align}
with $\mu,\nu =1,2,\dots,\mathcal{D}$ and $i,j=1,2,\dots,N$. The evolution of expectation values of GGMs can now be computed as
\begin{align}
\mi\frac{d}{dt} \langle\Lambda^i_\mu\rangle&= 
\langle [\Lambda^i_\mu,\hat H_\Lambda]\rangle\nonumber\\
   &=\sum_\mu v^i_\nu \langle [\Lambda^i_\mu,\Lambda^i_\nu]\rangle+\sum_{\sigma,j\neq i,\nu}w^{ij}_{\sigma,\nu}\langle[\Lambda^i_\mu,\Lambda^i_\sigma \Lambda^j_\nu] \rangle\label{eq:dLam}.
\end{align}
For each GGM $\Lambda^i_\mu$ we define a corresponding real phase space variable, $\lambda^i_\mu$. As in the ordinary TWA we will assume that the dynamics can be determined from statistical averages over trajectories of the phase space variables only. Thus in analogy to the case of $3$ spin-variables discussed after Eq.~\eqref{eq:heis}, our classical equations of motion follow from assuming a factorization between different sites $\langle \Lambda^i_\mu\Lambda^j_\nu\dots\Lambda^k_\sigma\rangle=\langle\Lambda^i_\mu\rangle\langle\Lambda^j_\nu\rangle\dots\langle\Lambda^k_\sigma\rangle$ for any nonequal site-index $i,j,\dots,k$ in Eq.~\eqref{eq:dLam}, and by replacing $\lambda^i_\mu(t) \equiv \langle \Lambda_\mu^i \rangle$.  

To define a phase space probability distribution for the inital state, we decompose each $\Lambda^i_\mu$  via its eigenvectors $\ket{a^i_\mu}$ with corresponding eigenvalues $a^i_\mu$, $\Lambda^i_\mu=\sum_{a^i_\mu}a^i_\mu\ket{a^i_\mu}\bra{a^i_\mu}$. As in spin-1/2 systems, where the Pauli matrices $\sigma_{x,y,z}$ can be measured to be $\pm 1$ in a projective measurement, the $a^i_\mu$ correspond to possible measurement outcomes of $\Lambda^i_\mu$.  We will focus on the experimentally relevant case of initial product states $\hat \rho = \prod_i \hat \rho^i$. Then, for each $\hat \rho^i$ we can define a corresponding probability distribution for our phase space variables, which we limit to a discrete set of values
\begin{align}
&~~\lambda^i_{\mu} \in \{a^i_\mu\} \quad \text{with probabilities}\nonumber \\&\quad p_{\mu}^i(\lambda^i_{\mu} = a^i_\mu)={\rm Tr}\left(\hat \rho^i\ket{a^i_\mu}\bra{a^i_\mu}\right). \label{eq:ddist}
\end{align}
The overall distribution factorizes for different variables on the same site and between sites, such that the probability for a configuration of all $\lambda^i_{\mu}$ being a certain combination of the eigenvalues $a_\mu^i$ for $i=1,2,\dots, N$ and $\mu=1, 2, ..., \mathcal{D}-1$ is given by $p(\{\lambda^i_{\mu} = a^i_\mu\})=\prod_{i,\mu} p_{\mu}^i(\lambda^i_{\mu} = a^i_\mu)$.

From the distribution \eqref{eq:ddist}, we can compute the expectation value of any observable on a fixed site $m$ via the expansion \eqref{eq:ggm_exp},
\begin{align}
  \langle \hat O^m\rangle 
 & = \sum_\mu c_\mu^m \langle \Lambda_\mu^m \rangle\nonumber\\
  &= \sum_\mu c_\mu^m \sum_{a_{\mu}^m} p_\mu^m(\lambda_\mu^m = a_\mu^m) \lambda_{\mu}^m = \sum_\mu c_\mu^m \llangle \lambda_\mu^m \rrangle.
  \label{eq:localcorr}
\end{align}
Here we defined the notation $\llangle \cdot \rrangle \equiv \sum_{\{a_\mu^i\}}p(\{\lambda^i_{\mu} = a^i_\mu\}) (\cdot)$, denoting the statistical average over any combination of eigenvalues for the $\lambda_\mu^i$ inside the brackets.
Note that for a Gaussian distribution for three spin-variables it is not possible to exactly represent any observable on the Hilbert space of a single-spin, since the three spin-operators on a site $m$, $\hat S^m_{x,y,z}$ and $\id$ do not form a complete local basis for any operator. Thus, for example the observable $(\hat S_z^m)^4$ cannot be expanded as linear superposition of the matrices $\{\hat S^m_{x,y,z},\id\}$ and has to be evaluated as 4-th order moment from the probability distribution. As a consequence, for a spin coherent state polarized along $x$, the Gaussian distribution does not reproduce $(\hat S_z^m)^4$ correctly (see~\ref{app:gaussian_tom}). 

For our GDTWA scheme, we use a discrete configuration selected from the distribution \eqref{eq:ddist} as initial condition for a classical evolution of the phase-space variables $\lambda^{i}_{\mu}(0) \in \{a_\mu^i\}$ for all $\mu$ and $i$. This configuration is then numerically evolved according to the equations of motion that we derive from the factorization of Eq.~\eqref{eq:dLam}. Those equations are in general non-linear and can be written in the form $d\bm{\lambda}(t)/dt = \mathcal{M}(\bm{\lambda}(t)) \bm{\lambda}(t)$. Here $\bm{\lambda}$ is a vector of all $N \times (\mathcal D-1)$ phase-space variables $\lambda_\mu^i$, and the matrix $\mathcal{M}(\bm{\lambda}(t))$ depends in general on the configuration at time $t$. We solve those equations numerically for the selected initial condition and obtain a mean-field trajectory, $\lambda^{m}_{\nu}(t)$. We repeat this procedure and average over the possible initial configurations. We thus obtain e.g.~an approximation for the expectation value of a GGM $\Lambda_\nu^m$ at time $t$ as $\langle \Lambda_\nu^m \rangle(t) \approx\llangle \lambda^{m}_{\nu}(t) \rrangle $.

If the problem is linear, i.e.~if $\mathcal{M}(\bm{\lambda}(t))=\mathcal{M}$ is time-independent, e.g.~if there are only single-spin terms in Eq.~\eqref{eq:dLam}, then the time-dependent solution in our scheme is  given by $\langle \Lambda_\nu^m \rangle(t) = \exp(\mathcal{M}t)\llangle \lambda^{m}_{\nu}(0) \rrangle $, and since $\llangle \lambda^{m}_{\nu}(0) \rrangle = \langle \Lambda^{m}_{\nu}\rangle(0)$ via Eq.~\eqref{eq:localcorr}, the time-evolution scheme is exact. If the problem is non-linear, which is generally the case in presence of inter-spin interaction terms in Eq.~\eqref{eq:dLam}, then the statistical average $\llangle\lambda^{m}_{\nu}(t) \rrangle$ will also depend on intra-spin correlations in the initial state, also within a single spin, e.g.~$\llangle\lambda^i_\mu (0) \lambda_\nu^i (0)\rrangle$, and higher orders. In general, those correlations do not coincide with the exact correlations obtained from the true quantum state. For example, it is straightforward to see that per definition of our discrete probability distribution, all correlations factorize between different GGMs on the {\em same} site, e.g.~$\llangle{\lambda^i_\mu(0) \lambda_\nu^i}(0)\rrangle =\llangle{\lambda^i_\mu(0) }\rrangle\llangle{\lambda_\nu^i(0)}\rrangle $ with $\mu \neq \nu$. This is not necessarily  true for the quantum mechanical intra-spin correlations of a generic single-spin state. For example, for some states $\hat \rho^i_g$, ${\rm Tr}[\hat \rho^i_g (\Lambda^i_\mu \Lambda^i_\nu +\Lambda^i_\nu \Lambda^i_\mu)/2] \neq {\rm Tr}(\hat \rho^i_g \Lambda^i_\mu) {\rm Tr} (\hat \rho^i_g  \Lambda^i_\mu)$, where we have to use a symmetrization since the GGMs do not generally commute. 

In~\ref{app:diagonal_state_exact} we show, that for ``diagonal'' product states all initial intra-spin correlations can be perfectly reproduced by our distribution \eqref{eq:ddist}. We define those states as $\hat  \rho = \prod_i \ket{\alpha_0}_i\bra{\alpha_0}$ where $\alpha_0 = 1,2,\dots,\mathcal{N}$, and $\ket{\alpha_0}_i$ is any of the single-spin eigenstates of the matrices defined in Eqs.~\eqref{eq:Ds}. It is important to mention that  any product of non-diagonal pure states can be brought into the form $\hat  \rho = \prod_i \ket{\alpha_0}_i\bra{\alpha_0}$, via local unitary transformations. Therefore, the diagonal assumption is not an actual restriction and in our simulation scheme we can exactly describe generic products of initial pure states. This can be achieved by applying the same unitary transformations to the equations of motion (see~\ref{app:diagonal_state_exact}). Note that, alternatively, one may also use a Gaussian approximation for the distribution of the initial $\lambda_\mu^i$~\cite{anatolicluster2018, Davidson2015}, i.e.~for all $\mathcal{D}-1$ generalized Bloch sphere variables.  However, we point out that in contrast to the discrete distribution given by  Eq.~\eqref{eq:ddist},  the Gaussian sampling not only  does not reproduce all initial intra-spin correlations correctly, but also can lead to worse longer-time predictions. For example, the dynamics of collapses and revivals has been only observed with the exact discrete sampling~\cite{Schachenmayer2015a}.

To compute the time-dependent expectation value of an arbitrary multi-spin observable, $\hat{\mathcal{O}}$, we expand
\begin{align}
\hat{\mathcal{O}}
  &= \sum_{\mu_1,\dots,\mu_N} C_{\mu_1,\mu_2,\dots,\mu_N} \Lambda_{\mu_1}^1  \Lambda_{\mu_2}^2   \dots   \Lambda_{\mu_N}^N 
  \label{eq:gen_op_exp}
\end{align}
and approximate
\begin{align}
    \langle \hat{\mathcal{O}}\rangle(t) 
  \approx  
    \sum_{\mu_1,\mu_2,\dots,\mu_N} C_{\mu_1,\mu_2,\dots,\mu_N}  \llangle\prod_{i}  \lambda_{\mu_i}^i (t)\rrangle.
  \label{eq:gdtwa}
\end{align}
This equation is a discrete analog of the TWA method from Eq.~\eqref{eq:TWA_gaussian} for a phase space extended to a high-dimensional generalized Bloch sphere. Approximating Eq.~\eqref{eq:gdtwa} with a finite number of samples, $n_s$, from the discrete probability distribution is what we denote as generalized discrete truncated Wigner approximation (GDTWA). The key idea underlying the method is also illustrated in Fig.~\ref{fig:scheme}.

As a clarifying example, let's also explicitly provide a formula for computing the evolution of a single-spin observable, $\hat O^m$. After selecting a single random configuration for all phase space variables $\{\lambda_{\mu}^{i}\}$ according to $p_\mu^i(\{\lambda_\mu^i = a_\mu^i\})$, we denote the subsequent classical evolution of the variables at site $m$ for this sample ``$s$'' as $\lambda_{\mu, {\rm cl}}^{m,[s]} (t)$. Then, the evolution of $\hat O^m$ is computed as
\begin{align}
\langle \hat O^m \rangle (t) &\approx \frac{1}{n_s} \sum_{s}^{n_s} \sum_\mu \lambda_{\mu, \rm{cl}}^{m,[s]} (t) {\rm Tr}[\hat O^m \Lambda_{\mu}^m].\label{eq:19}
\end{align}
Note that in contrast to a continuous Wigner function approach, here in principle only a finite number of numerical samples $n_s$ is needed to compute Eq.~\eqref{eq:gdtwa}. In practice, however, this number increases exponentially with the $N$. The $n_s$ needed to obtain converged results in Eq.~(\ref{eq:19}) depends on the system size and the observable. In practice for typical realistic problems (with hundreds of spins as used in Sec.~\ref{sec:ent}) we find a number on the order of a few $10^4$ samples sufficient. Note that we find that for collective observables the number of samples necessary for convergence typically decreases with $N$.

To further explain the main idea behind  this approach and its capabilities, we first consider as an example an array of spin $S=1/2$ particles. In this case, the three nontrivial $\Lambda_\mu$ for each spin are proportional to standard Pauli matrices, $ \Lambda_{x,y,z}=\sigma_{x,y,z}/\sqrt{2}$, each of which has eigenvalues $\pm 1/\sqrt{2}$,  and  eigenstates described by states aligned (anti-aligned) along the corresponding direction: $\ket{\uparrow_{x,y,z}(\downarrow_{x,y,z})}$. For a spin initially polarized as $\ket{\uparrow_z}$, the initial values of $\lambda_{x}$ in phase space take its two eigenvalues, $\pm 1/\sqrt{2}$, with equal probability 
${\rm Tr}
[\ket{\uparrow_z}
\braket{\uparrow_z|\uparrow_x}
\bra{\uparrow_x}] 
= |\braket{\uparrow_z|\uparrow_x}|^2 = |\braket{\uparrow_z|\downarrow_x}|^2=0.5$. 
In the same manner probabilities for $\lambda_y=\pm 1/\sqrt{2}$ are $0.5$, while $\lambda_z=1/\sqrt{2}$ is fixed (probability 1). Our GDTWA prescription thus reduces to the DTWA sampling introduced in Ref.~\cite{Schachenmayer2015a}.

For $S=1$, there are three states $\ket{+1},\ket{0},\ket{-1}$ for each spin, and $\Lambda_\mu$ consists of the identity matrix and 8 nontrivial Gell-Mann matrices:

\begin{align}
\Lambda_1&=
\begin{pmatrix}
 0 & \frac{1}{\sqrt{2}} & 0 \\
 \frac{1}{\sqrt{2}} & 0 & 0 \\
 0 & 0 & 0 \\
\end{pmatrix}
&
\Lambda_2&=
\begin{pmatrix}
 0 & 0 & \frac{1}{\sqrt{2}} \\
 0 & 0 & 0 \\
 \frac{1}{\sqrt{2}} & 0 & 0 \\
\end{pmatrix}
\nonumber\\
\Lambda_3&=
\begin{pmatrix}
 0 & 0 & 0 \\
 0 & 0 & \frac{1}{\sqrt{2}} \\
 0 & \frac{1}{\sqrt{2}} & 0 \\
\end{pmatrix}
&
\Lambda_4&=
\begin{pmatrix}
 0 & -\frac{i}{\sqrt{2}} & 0 \\
 \frac{i}{\sqrt{2}} & 0 & 0 \\
 0 & 0 & 0 \\
\end{pmatrix}
\nonumber\\
\Lambda_5&=
\begin{pmatrix}
 0 & 0 & -\frac{i}{\sqrt{2}} \\
 0 & 0 & 0 \\
 \frac{i}{\sqrt{2}} & 0 & 0 \\
\end{pmatrix}
&
\Lambda_6&=
\begin{pmatrix}
 0 & 0 & 0 \\
 0 & 0 & -\frac{i}{\sqrt{2}} \\
 0 & \frac{i}{\sqrt{2}} & 0 \\
\end{pmatrix}
\nonumber\\
\Lambda_7&=
\begin{pmatrix}
 \frac{1}{\sqrt{2}} & 0 & 0 \\
 0 & -\frac{1}{\sqrt{2}} & 0 \\
 0 & 0 & 0 \\
\end{pmatrix}
&
\Lambda_8&=
\begin{pmatrix}
 \frac{1}{\sqrt{6}} & 0 & 0 \\
 0 & \frac{1}{\sqrt{6}} & 0 \\
 0 & 0 & -\sqrt{\frac{2}{3}} \\
\end{pmatrix}
.
&
\end{align}
An initially polarized state $\ket{+1}$, can now be represented via probabilities for eigenvalues of $\Lambda_{1,2,...8}$. Specifically, with $1/2$ probability one chooses  $\lambda_{1},\lambda_{2},\lambda_{4},\lambda_{5}$ at value $1/\sqrt{2}$, and with $1/2$ probability one chooses the value $-1/\sqrt{2}$. In the following, we denote this as  
$\lambda_{1,2,4,5}=\{-1/\sqrt{2},1/\sqrt{2}\}$,  
with probability $p_{1,2,4,5}=\{0.5,0.5\}$. 
Then, further we have fixed 
$\lambda_{3,6}=0$ with $p_{3,6}=1$,  $\lambda_{7}=1/\sqrt{2}$ with $p_{7}=1$, and $\lambda_{8}=1/\sqrt{6}$  with $p_{8}=1$ (See Fig.~\ref{fig:scheme} for an illustration of this distribution).

While the  initial states discussed above are still spin coherent states, the state $\ket{0}$, however, is an example of a state that cannot be  represented with the conventional Gaussian TWA approach. In the GDTWA it can be easily simulated with $\lambda_{1,3,4,6}=\{-1/\sqrt{2},1/\sqrt{2}\}$, $p_{1,3,4,6}=\{0.5,0.5\}$,  $\lambda_{2,5}=0$, $p_{2,5}=1$,  $\lambda_{7}=-1/\sqrt{2}$, $p_{7}=1$, and $\lambda_{8}=1/\sqrt{6}$, $p_{8}=1$.

\section{Comparison with exact diagonalization}
\label{sec:benchmark}

\begin{figure*}
\centering
\includegraphics[width=0.98\textwidth]{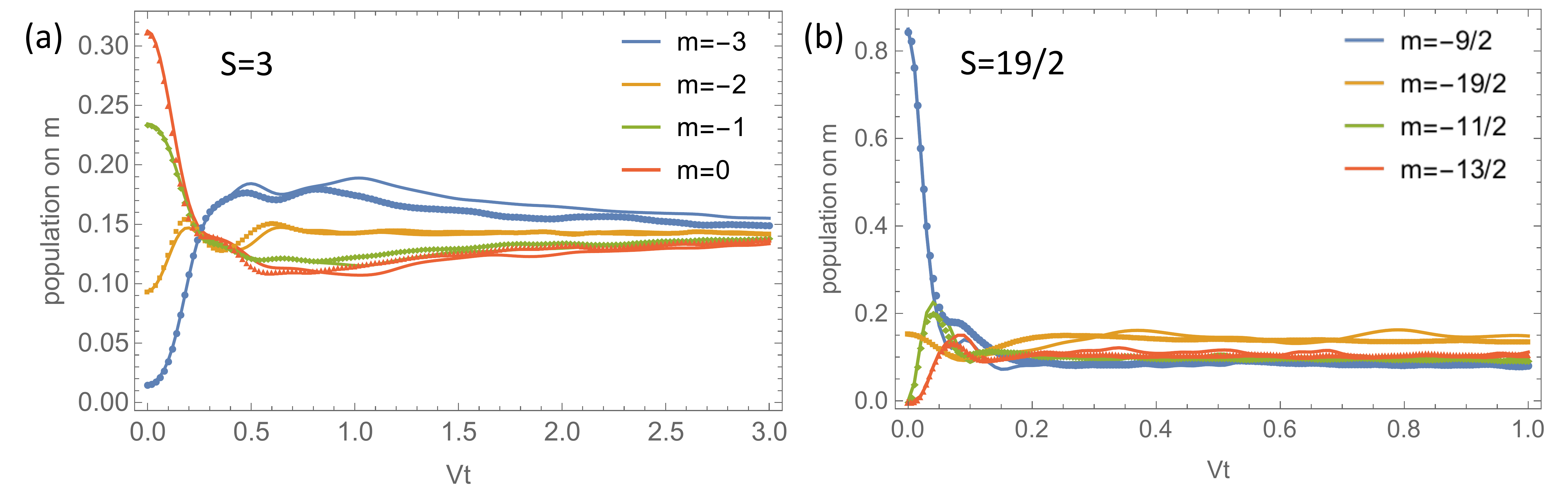}
\caption{\label{fig:ed}  Dynamics in a spin model with dipolar long-range interactions computed from ED (solid lines) and GDTWA (dots). Shown are the evolutions of the averaged populations in different Zeeman levels. (a) Spin-3 (Cr) atoms on a $2\times 2\times 2$ cubic lattice. The initial state is a spin coherent state polarized along $x$ ($\theta = \pi/2$, quantization axis along $z$). (b) Population dynamics in the four most populated Zeeman levels for a 1D chain of spin-19/2 (Er) atoms, with $\theta_{i,j}\equiv \pi/2$ and $N=5$. Here, the initial state of each atom $i$ is a coherent superposition of two Zeeman levels: $\ket{\psi}_i=\sqrt{0.85}\ket{m=-9/2}+\sqrt{0.15}\ket{m=-19/2}$.}
\end{figure*}

We now proceed to benchmark  the validity of the GDTWA by comparison with the dynamics obtained from numerical exact diagonalization (ED). Specifically, we will consider the dynamics induced by dipolar interactions, which are currently investigated in a variety of  experimental platforms, including polar molecules~\cite{Bohn2017}, magnetic atoms~\cite{Lepoutre_Out-of-_2019,manfred2019}, NV centers~\cite{Choi2017}, etc. The Hamiltonian for such systems can be written in the form of an anisotropic XXZ spin model:
\begin{eqnarray}
\hat H_{dd}&=&\frac{1}{2}\sum_{i,j\neq i}V_{i,j}[\hat S_z^i\hat S_z^j-\frac{1}{2}(\hat S_x^i\hat S_x^j+\hat S_y^i\hat S_y^j)],\label{eq:Hdip}
\end{eqnarray}
where $V_{i,j}=V_{dd}(1-3\cos^2\theta_{i,j})/r_{i,j}^3$ is the strength of the interaction between two atoms $i$ and $j$, whose distance is $r_{i,j}$ and $\theta_{i,j}$ is the angle between the vector ${\bf r}_{i,j}$ and the dipole orientation typically set by an external quantization field. For magnetic atoms such as chromium ($S=3$) and erbium ($S=19/2$ for fermions, and $S=6$ for bosons), $V_{dd}=\mu_0(\mu_Bg)^2/4\pi$, with $\mu_0$ the magnetic permeability of vacuum, $\mu_B$ the Bohr magneton, and $g$ the Land\'{e} g-factor. For convenience, we define a bare dipolar interaction strength $V=V_{dd}/d^3$, with $d$ the smallest  spacing between different lattice sites.

{\it Spin-level Populations: }
An appealing property of  magnetic atoms  is their sizable spin, which by itself is responsible for enhanced dipolar interactions.  A large spin provides a great degree of tunability through the control of  the various  multiple internal spin states. In recent  experiments  the spin dynamics has been activated by preparing  two types of initial states: i) a spin coherent state of atoms~\cite{Yan2013,Lepoutre_Out-of-_2019} $\ket{\psi(t=0)}=\otimes_j \exp({\mi\theta\hat S_y^j})\ket{-S}_j$, with $\theta\in[0,\pi]$ the tipping angle, or ii) all atoms in the same Zeeman level \cite{dePazMISF,depaz2013,depaz2016,manfred2019} $\ket{\psi(t=0)}=\otimes_j\ket{m_0}_j$, where $m_0=-S,-S+1,\dots,S,$ and $j=1,2,\dots,N$ labels atoms (see also~\ref{app:diagonal_state_exact} for discussions on initial states in GDTWA). In Fig.~\ref{fig:ed},  we compute the evolution of population in different Zeeman levels $m$ averaged over all spins, $n_{m}(t) = (1/N) \sum_j |\braket{\psi(t)|{m}}_j|^2$, starting from the different initial states for both scenarios with $S=3$ and $S=19/2$. We compare a simulation using the GDTWA with the exact ED result. The very good agreement between GDTWA and ED shows that the GDTWA captures the  dynamics for all Zeeman levels well under dipolar exchange. While, as expected the GDTWA provides quantitatively exact results for short times, remarkably it features also qualitatively the long-time behavior well. Note that an attempt to simulate the population dynamics of Fig.~\ref{fig:ed} using a TWA approach with Gaussian sampling of three spin-variables leads to unphysical results as shown in \ref{app:gaussian_tom}.

\begin{figure*}
\centering
\includegraphics[width=0.98\textwidth]{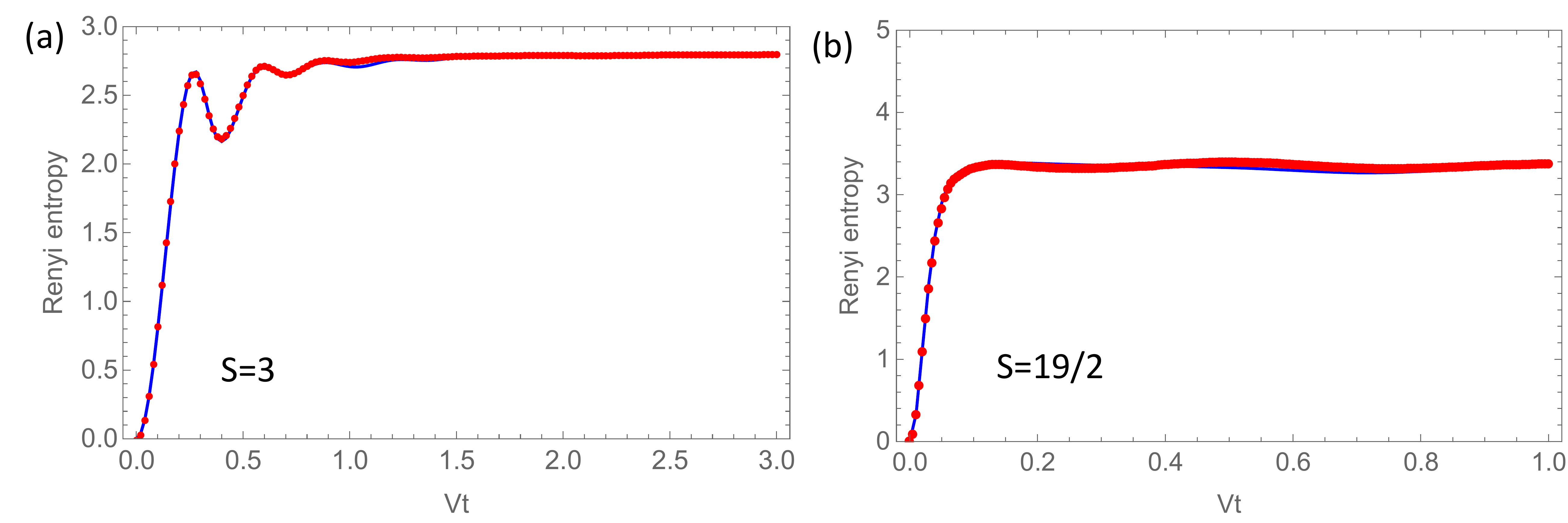}
\caption{\label{fig:edrenyi}  Evolution of the second Renyi entropy of a single spin density matrix $\hat \rho_c$ computed with ED (solid lines) and GDTWA (dots), corresponding to the population dynamics shown in Fig.~\ref{fig:ed}. (a) $S=3$, and $\hat\rho_c$ corresponds to a spin on the edge of the $2\times 2 \times 2$ cube. (b) $S=19/2$, and $\hat \rho_c$ is for the central spin in the chain. The initial conditions are the same as those used in Fig.~\ref{fig:ed} for $S=3$ and $S=19/2$, respectively.}
\end{figure*}

{\it Single-spin Density Matrix: } In addition to Zeeman level populations and spin projections $\langle \hat S^i_{x,y,z} \rangle$, the GDTWA also provides access to all other elements of any single-spin density matrix element. It is important to note that this cannot be obtained with a conventional Gaussian sampling. In particular, while in principle a state-tomography can also be performed from combinations of expectation values of powers of spin operators, the Gaussian sampling does not properly account for high order moments. For example, in the case of a spin coherent state along the $z$-axis the Gaussian sampling provides incorrect results for expectation values $\langle \hat S_{x,y}^n \rangle$ with $n>3$. This implies that for models with $S>3/2$, a state tomography leads to significant errors for density matrix elements, already in the initial state (see~\ref{app:gaussian_tom}). In the GDTWA, the ability to predict all density matrix elements allows us to compute entanglement measures such as  the Renyi entropy for a single spin in the coupled system, $\mathcal{S}^c_\alpha=\frac{1}{1-\alpha}{\log_2}{\rm Tr}[\hat\rho_c^\alpha]$, where $\hat \rho_c$ is the reduced density matrix of a single spin $c$, and $\alpha\geq 0$.  In Fig.~\ref{fig:edrenyi}, we compare the evolution of the  second  Renyi entropy ($\alpha=2$) computed with GDTWA and ED for the same scenarios as in Fig.~\ref{fig:ed}. For both initial states, with $S=3$ and $S=19/2$ respectively, we find excellent agreement for all timescales. Note that in~\ref{app:benchmarktheta}, we provide further comparisons for type i) initial state with different $\theta$ to better analyze the validity of the GDTWA. There we observe that the GDTWA can provide nearly exact predictions, especially  in cases  where the system quickly reaches a steady state with high local entropy. Nevertheless, deviations from the exact Renyi entropy evolution are seen  at low angles ($\theta \to 0$). This is an interesting observation that needs further investigation  given that   the $\theta \to 0$ limit is where a simple  mean-field treatment is enough to accurately reproduce the short time dynamics.  

\begin{figure*}
\centering
\includegraphics[width=0.5\textwidth]{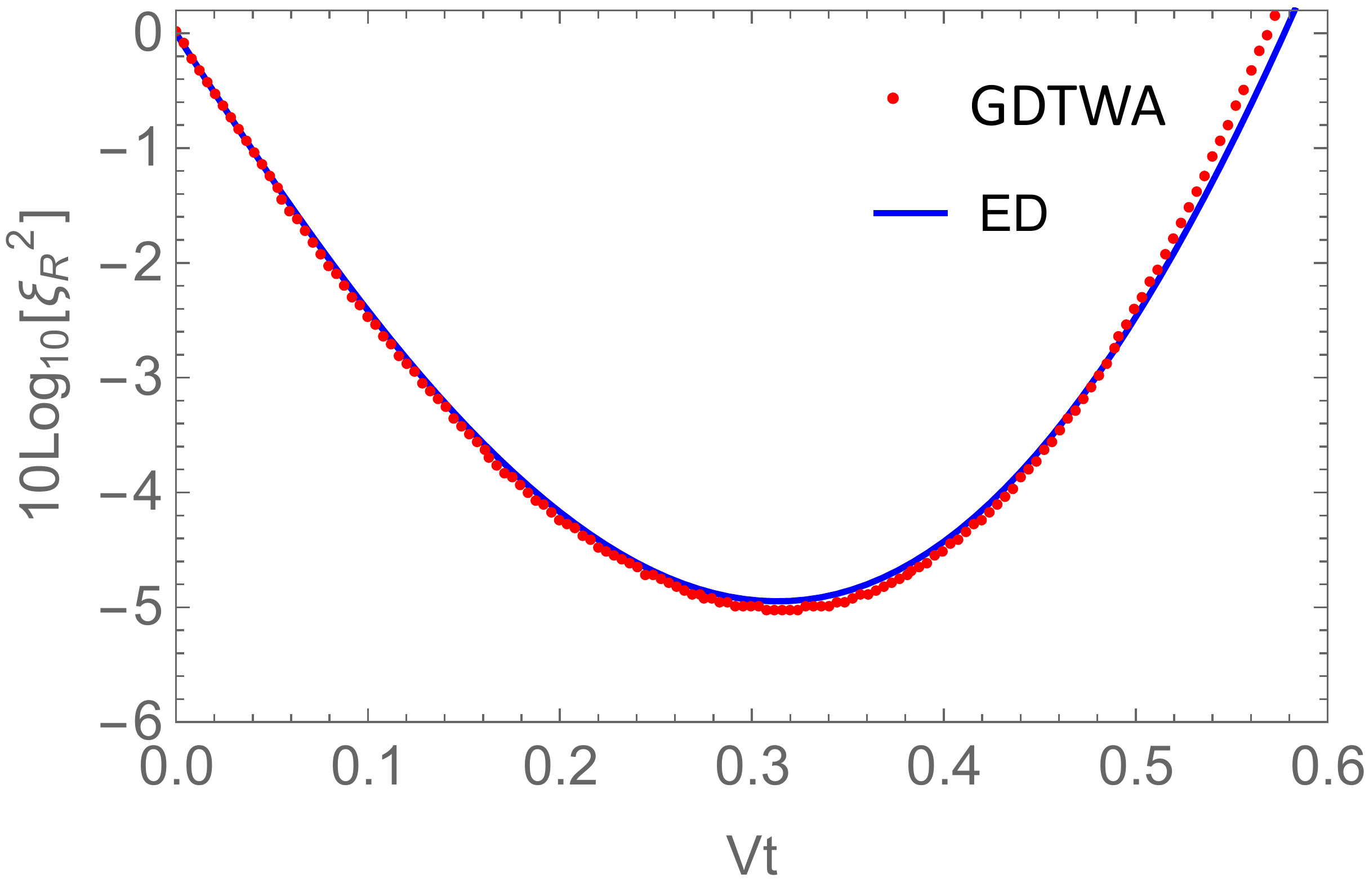}
\caption{\label{fig:squeezbmk}Evolution of the quantum spin squeezing parameter $\xi_R^2$ from Eq.~\eqref{eq:xi_squeeze} calculated with ED (solid line) and GDTWA (dots). The system is a 1D chain with $N=8$ atoms with spin $S=2$ and with dipolar interactions (dipole orientation perpendicular to the chain). The initial state is a spin coherent state polarized along the chain direction, i.e.~of type i) with $\theta=\pi/2$. The GDTWA captures the interaction induced spin-squeezing ($\xi_R<1$) nearly perfectly.}
\end{figure*}

{\it Spin Squeezing:} In a  many-body system,  interactions can generate quantum spin squeezing, which can fundamentally improve measurement precision~\cite{uedasqu,masqu2011,qumetrreview2018}. Spin squeezing  is of interest to many current experiments~\cite{squbec2008,vuletic2010,Bohnet2016,KasevichNature2016}. Here we demonstrate that the GDTWA is also applicable for studying such a phenomenon. In particular,  we study the spin squeezing parameter first introduced by Wineland et al.~as a measure of phase sensitivity~\cite{winelandxi,winelandxi2},
\begin{align}
\label{eq:xi_squeeze}
\xi_R^2=\frac{2NS{\rm min}[\Delta S_\perp^2]}{|\vec{S}|^2},
\end{align}
where $\Delta S_\perp^2=\langle \hat S_\perp^2\rangle-\langle\hat S_\perp\rangle^2$ is the variance measured perpendicular to the collective spin $\vec{S}$ direction,  and $|\vec{S}|=\sqrt{\langle\hat S^x\rangle^2+\langle\hat S^y\rangle^2+\langle\hat S^z\rangle^2}$ is the length of the collective spin. As indicated by this definition, the spin squeezing parameter involves quantum correlations among many atoms. For a spin coherent state, $\Delta S_{\perp}^2=NS/2$, $\xi_R^2=1$. On the other hand $\xi_R^2<1$ implies reduced phase noise. As shown in Fig.~\ref{fig:squeezbmk}, the result obtained with GDTWA again agrees almost perfectly with those obtained from ED. Note that compared to Eq.~\eqref{eq:gdtwa}, here we use a slightly modified method to compute collective spin-observables such as $\langle (\hat S^x)^2\rangle$. In various benchmark situations we find that this modified procedure provides much more precise predictions for the spin-squeezing parameter than Eq.~\eqref{eq:gdtwa}. We discuss this in \ref{app:squeez}.

\section{\label{sec:ent}Spin dynamics in  many-body systems}

While the system sizes accessible with  ED are  remarkably small for large $S$, the GDTWA enables us to overcome this limitation and to perform  investigations of spin dynamics in  large ensembles with large $S$. This capability  is very important  to 
quantitatively perform  comparisons to experiments, in particular  in the presence of collective behaviors, where  finite size effects matter. In the following sections we will employ the GDTWA to obtain new insight for those systems by studying: time-dependent re-distribution of populations between Zeeman levels and resulting quantum thermalization for this observable, the evolution of entanglement among the dipoles, the evolution of the contrast of spin coherence, and spin-squeezing.

\subsection{Spreading of Zeeman level populations}

\begin{figure*}
\centering
\includegraphics[width=1\textwidth]{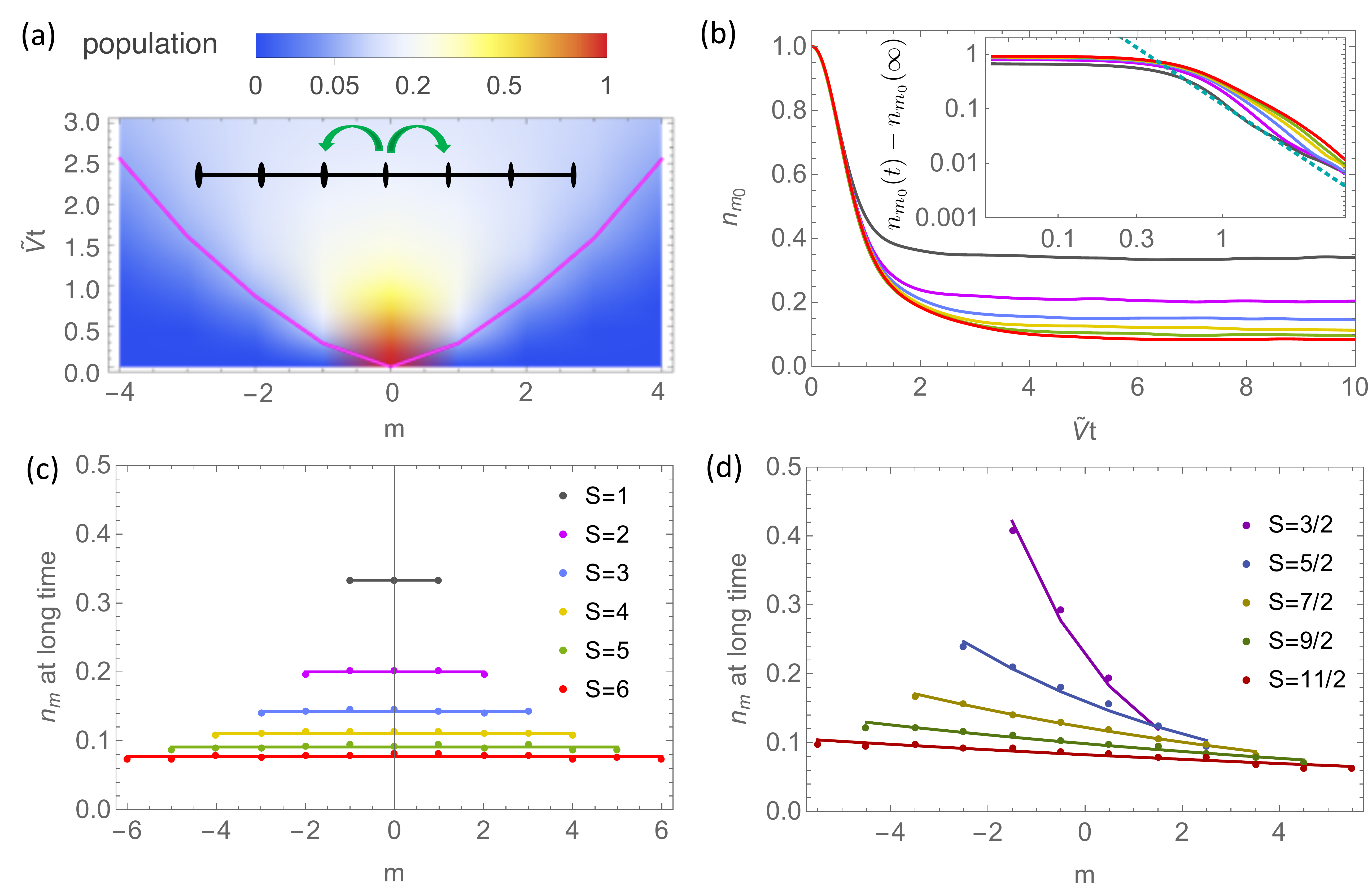}
\caption{\label{fig:thermS} (a) Redistribution of population among Zeeman levels for $S=3$. The magenta line denotes a contour at $0.05$. The inset illustrates the propagation of population to adjacent internal levels, due to the dipolar exchange term $\hat S_i^+\hat S_j^-$.  (b) Dynamics of population on the initial Zeeman level $m_0=0$, under dipolar interactions for various atomic species with increasing $S=1,\dots,6$ (from top to bottom). The inset shows the dynamics in log-log scale, with the vertical axis showing the offset from $1/(2S+1)$, i.e., the thermal prediction. The cyan dotted line shows the scaling $\propto 1/t^2$. (c) (d) Long time population distribution on $2S+1$ atomic states for integer and half-integer $S$, respectively, where the solid lines indicate the prediction from a thermal ensemble, and the dots are the long time values calculated with GDTWA. For all panels we use a 3D lattice of size $6\times 6\times 6$ and dipoles are aligned along $z$. }
\end{figure*}

Here, we apply the GDTWA approach to study the spin dynamics in a 3D lattice of dipoles for various $S$, with the interaction Hamiltonian given by Eq.~(\ref{eq:Hdip}). We prepare the system in the type ii) initial state, which is not an eigenstate of the many-body Hamiltonian $H_{dd}$, and thus the population of the initial Zeeman level will generally change with time. Specifically, the dipolar exchange will lead to a redistribution of population over all $2S+1$ levels (see Fig.~\ref{fig:thermS}(a)). As demonstrated in Ref.~\cite{manfred2019}, the effective dipolar exchange strength given by  $V\gamma(S,m_0)$ can be used as the  characteristic energy scale of the spin dynamics, with  $\gamma(S,m_0)$ given by 
\begin{align}
\gamma(S,m_0)&=\sqrt{f_+f_-(f_++1)(f_-+1)},\\
f_{\pm}&=S\pm m_0,\nonumber
\end{align}
where $m_0$ is the initially populated spin level. Correspondingly we define $\tilde{V}=V\gamma(S,m_0)$. As shown in Fig.~\ref{fig:thermS}(b), for various values of $S$, the initial dynamics happens at a characteristic rate $1/\tilde{V}$. 

It is worth to note that for these initial conditions within a mean-field approximation one observes  the spin-level populations to remain  fixed at the initial value without any evolution. Quantum fluctuations of the initial state are what drives the dynamics in this case  and therefore need to be accounted for. The GDTWA includes those initial fluctuations in an exact way and thus can describe the evolution of the level populations. Note that these simulations are done for a 3D system with $\sim 200$ atoms, which corresponds to a full Hilbert space $\sim 10^{240}$ states for $S=6$ and is far beyond the capacity of ED.

At long times, when the system  thermalizes, the spin dynamics shows saturation to different values for different $S$. According to the theory of closed system thermalization, given e.g.~by the Eigenstates Thermalization Hypothesis (ETH), it is expected that for a non-integrable system far away from a localized phase~\cite{DAlessio_Fromqu_2016, Deutsch_Eigenst_2018, Basko_Metali_2006, Nandkishore_Many-Bo_2015}, at long times the expectation values of local observables can be effectively described by a thermal equilibrium state, $\hat\rho_{\infty}={\rm exp}(-\beta\hat H_T-\mu\hat S^z)/\Xi$, with $\Xi={\rm Tr}[{\rm exp}(-\beta\hat H_T-\mu\hat S^z)]$, $\hat H_T=\hat H_{dd}$, and $\hat S^z=\sum_i\hat S_z^i$. The effective inverse temperature $\beta$ and chemical potential $\mu$ are set by the total energy and magnetization, respectively, which are the  conserved quantities in our problem. Those are determined by the initial state and conserved under the  unitary evolution. In our system, the  total energy for the initial state is $E_0=\sum_{i, j\neq i}V_{i,j}m_0^2/2$. In a high temperature limit $\beta \to 0$, energy conservation leads to
\begin{align}
  \beta\approx -\frac{24E_0}{S^2(S+1)^2\sum_{i,j\neq i}V_{i,j}^2}.
\end{align}
For a 3D lattice where dipolar interactions vary in  sign, $E_0\approx 0$ and thus we take $\beta=0$ in the following. Then $\mu$ can be easily determined from $\sum_{m=-S}^S m {\rm exp}(-\mu m)/\Xi=m_0$, with $\Xi=\sum_{m=-S}^S {\rm exp}(-\mu m)$, and we can obtain the equilibrium population  in a spin $m$ state:
\begin{align}
n_{m}(\infty)=\frac{{\rm Tr}[\hat \rho_{\infty}\hat n_{m}]}{\Xi}.
\end{align}
As shown in  Fig.~\ref{fig:thermS} (c) and (d), the long-time populations obtained from the GDTWA agree extremely well with the quantum thermalization predictions for both integer and half-integer $S$. For integer $S$, an initial state with $m_0=0$ results in equal population on all Zeeman levels in equilibrium. This is not the case for half-integer spins where the steady state populations are not equal. The different steady-state  behavior seen in  integer and half-integer spin systems is  reminiscent of their different  low energy spectra at  equilibrium. 

The agreement shown here suggests that in the long-time limit GDTWA simulations for simple single-spin observables approach the results expected from closed system quantum thermalization, an effect also observed previously for different initial states for $S=1/2$~\cite{Acevedo_Explori_2017} and $S=3$~\cite{Lepoutre_Out-of-_2019} models. Note that while our simulations do not constitute a general proof of thermalization in the GDTWA, our results here suggest that in models which feature quantum thermalization, i.e.~in models which are  not  many-body localized, and  with local observables that  thermalize  quickly, the GDTWA can give accurate predictions for simple observables not only at short times, but over all relevant timescales.

Furthermore, the knowledge of the equilibrium values also allows us to use the GDTWA to investigate how such a quantum system approaches thermalization. Recent works have suggested a power-law relaxation in the long-time dynamics~\cite{Tavora_Inevit_2016, Tavora_Power-_2017, santos2017analytical,thermtimescale2019}. Interestingly, as illustrated in the inset of Fig.~\ref{fig:thermS} (b), here we observe a power-law decay towards  thermalization at long times, which roughly follows $n_{m_0}(t)-n_{m_0}(\infty)\sim 1/t^2$, with $n_{m_0}(t)$ the population on state $m_0$ at time $t$. 

\subsection{Entanglement build-up measured through the Renyi Entropy}

\begin{figure*}
\centering
\includegraphics[width=0.6\textwidth]{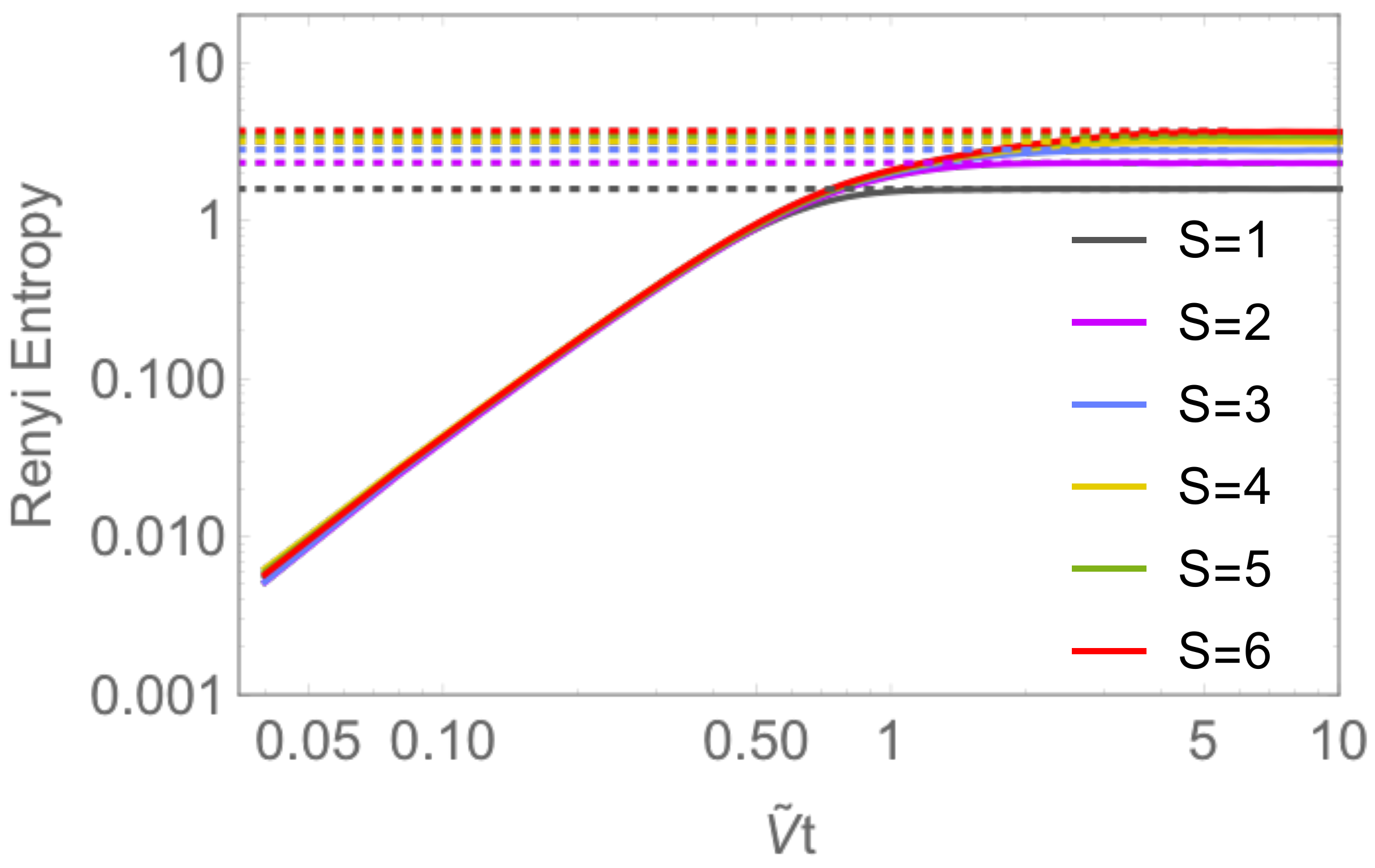}
\caption{\label{fig:renyi}  Evolution of second Renyi entropy for a single spin density martix $\hat \rho_c$ computed for the evolution of Fig.~\ref{fig:thermS}(b). For $\hat \rho_c$ we choose a central spin in the cube, and the dotted lines indicate the values of maximum entropy, i.e.~maximum bi-partite entanglement to the other spins.}
\end{figure*}

As mentioned in Sec.~\ref{sec:benchmark}, in GDTWA simulations we can easily compute the second Renyi entropy $\mathcal{S}^c_2=-\log_2{\rm Tr}[\hat\rho_c^2]$ for a subsystem of a single spin. The entropy  quantifies the purity of the subsystem and thus provides useful information on the bi-partite entanglement with all other spins, given that the state of the full system remains a pure state.  In Fig.~\ref{fig:renyi} we show the second Renyi entropy for a single spin during the spin dynamics of Fig.~\ref{fig:thermS}(b), with $\hat \rho_c$ chosen as the reduced density matrix of a central spin in the cube. Since our initial state is a pure state with $\mathcal{S}_2^c=0$, an increasing $\mathcal{S}_2^c(t)>0$ indicates a build-up of entanglement in the system. For all values of $S$, $\mathcal{S}^c_2$ increases quickly at short times in an algebraic way. It gradually approaches the maximum  second Renyi entropy of $\log_2{(2S+1)}$ which is allowed by the local Hilbert space size of $2S+1$, and which increases with $S$. This suggests that atoms with large $S$, rather than behaving more classically can feature richer entanglement-related quantum effects. Their increased local Hilbert space size allows larger entanglement entropies, and in our setup we find a quick approach to the maximally allowed Renyi entropy. Since for dipolar systems, such as the magnetic atoms discussed above, larger $S$ is associated with a larger net interaction strength, a larger spin also implies a faster generation of entanglement in the spin dynamics as confirmed in Fig.~\ref{fig:renyi}.

\subsection{Dynamics of contrast and intra-spin correlations}\label{sec:spcorr}

\begin{figure*}
\centering
\includegraphics[width=0.98\textwidth]{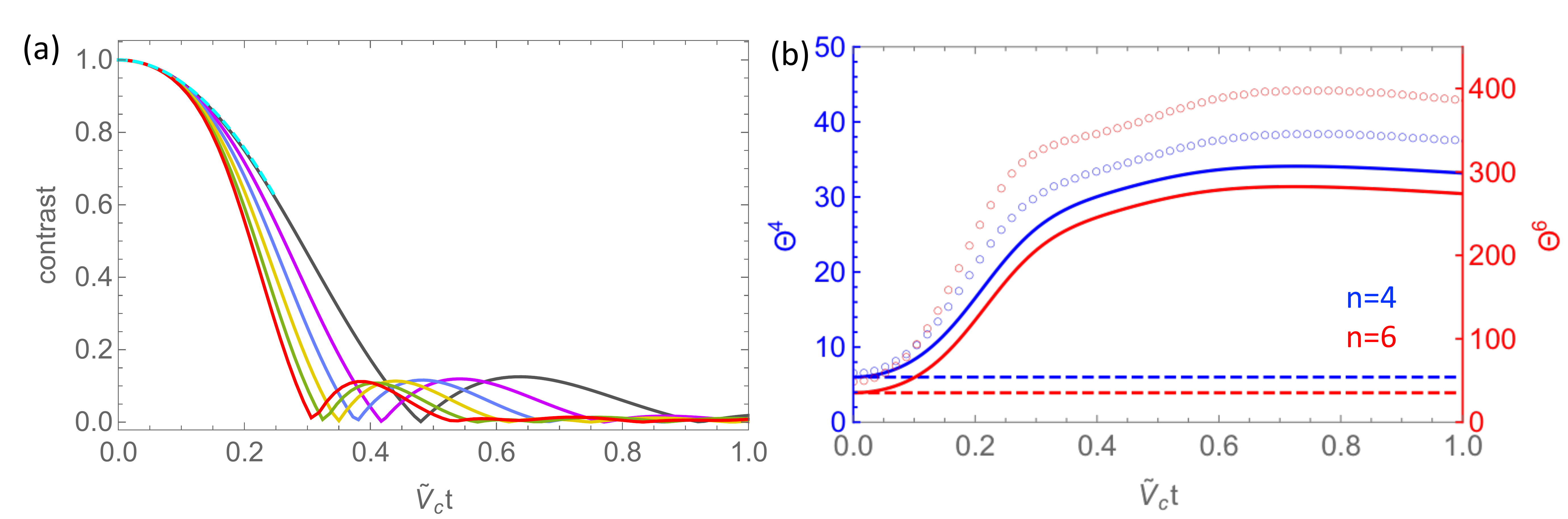}
\caption{\label{fig:spcorr} (a) Dynamics of  spin coherence contrast for various values of $S$.    The geometry of the lattice is the same as in Fig.~\ref{fig:thermS}, but the initial state is a spin coherent state of type i) with $\theta=\pi/2$. The cyan dashed line plots Eq.~(\ref{eq:csana}). (b) Comparison of GDTWA (solid lines) and Gaussian TWA (circles) simulations for the evolution of $\Theta_n=\langle\sum_j(\hat S_z^j)^n\rangle/N$, with $S=3$ and $n=4$ (blue) and $n=6$ (red). The dashed lines indicate the exact values of $\Theta_n$ at $t=0$ for a spin coherent state. For both (a) and (b) we use a 3D lattice of size $6\times 6\times 6$ and dipoles are aligned along $z$.     }
\end{figure*}

Another useful probe of many-body effects is the contrast of spin coherence, defined as $\mathcal{C}_{S}=\sqrt{\langle\hat S^x\rangle^2+\langle \hat S^y\rangle^2}/(NS)$, with $\hat S^{x,y,z}=\sum_{j}\hat S_j^{x,y,z}$. It has  been measured  via Ramsey spectroscopy in many experiments~\cite{Hazzard2014b,bishof2013,bromley2018}. Fig.~\ref{fig:spcorr}(a) shows  the evolution of $\mathcal{C}_{S}$ under the dipolar interactions in Hamiltonian~(\ref{eq:Hdip}), starting from the type i) spin coherent initial state with  tipping angle $\theta=\pi/2$.  Using a short-time expansion analysis, we find that the contrast initially decays as~\cite{Hazzard2014a} 
\begin{align}
\mathcal{C}_{S}(t)&\approx1-\frac{9S}{16N}\sum_{i,j\neq i}V_{i,j}^2t^2.\label{eq:csana}
\end{align}
Compared with the dynamics discussed in the previous section, where a larger spin $S$ can provide an enhanced dipole-dipole interaction $\propto\tilde{V}\propto S^2 V$, here the contrast decays slower than the timescale $\propto 1/\tilde{V}$.  Instead, the characteristic interaction strength for the contrast dynamics is $\tilde{V}_c=\sqrt{S}V$. This is seen in Fig.~\ref{fig:spcorr}(a), where as a function of $\tilde{V}_ct$ all lines collapse onto a single one for short times.

As mentioned in Sec.~\ref{sec:method}, such an initial spin state can also be represented in a conventional Gaussian TWA for three spin-variables. However, for $S>1/2$, quantum intra-spin correlations can develop~\cite{spinSsqu}. As discussed in Sec.~\ref{sec:method}, since in a Gaussian TWA only the spin  operators $\hat S_{x,y,z}^j$ are involved, such intra-spin correlations generally cannot be well captured. In the initial state, the GDTWA reproduces the exact results, differing from those of the Gaussian sampling, as also discussed in~\ref{app:gaussian_tom}. Furthermore, we find that the discrepancy between the Gaussian TWA and the GDTWA increases with time, as demonstrated in Fig.~\ref{fig:spcorr}(b). There, to illustrate this effect, we compare the evolution of $\langle\sum_j(\hat S_z^j)^n\rangle$ ($n=4,6$), computed with GDTWA and Gaussian TWA for three spin-variables. The Gaussian TWA provides incorrect initial values, and this  error  propagates over time.

\subsection{Quantum spin squeezing}
\begin{figure*}
\centering
\includegraphics[width=0.6\textwidth]{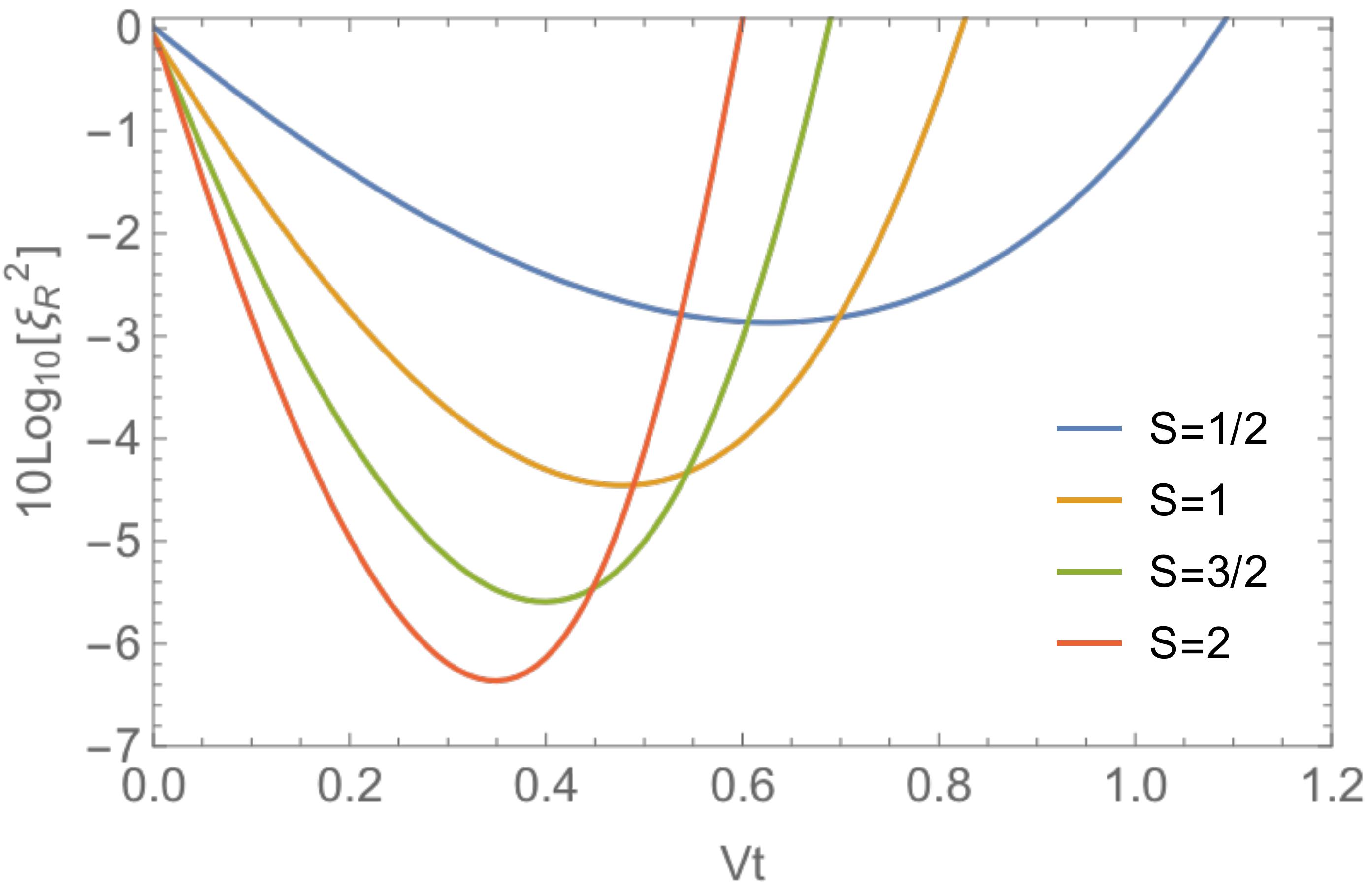}
\caption{\label{fig:squeez}Quantum spin squeezing parameter during the spin dynamics governed by $\hat H_{dd}$ and for varying $S$. To reduce finite-size effect, we consider a 1D chain with various large $N$ to adjust for a fixed $NS = 240$. Dipoles are aligned  perpendicular to the chain axis ($\theta_{i,j}\equiv\pi/2$). The initial state is a spin coherent state polarized along the chain direction, i.e.~of type i) with $\theta=\pi/2$.}
\end{figure*}

Lastly, to further characterize the quantum correlations generated during the dipolar spin dynamics, we study the spin squeezing parameter $\xi_R^2$ introduced in Sec.~\ref{sec:benchmark} for ensembles of spins with various $S$.  It is straightforward to show that $\xi_R^2\ge 1/(2NS)$. Hence, to make a fair comparison for systems of different spin $S$, in Fig.~\ref{fig:squeez}, we keep $NS$ constant and calculate  $\xi_R^2$ for various $S$. While  for $S>1/2$ spin  squeezing is no longer an entanglement witness given that part of the contributions come from intra-spin correlations~\cite{spinSsqu}, nevertheless the calculations do suggest an enhanced phase sensitivity potentially provided by large $S$ particles. Explicitly, compared to the conventional spin-1/2 particles, a system of $S=2$ shows significantly larger spin squeezing and also the timescale for reaching maximum squeezing is shortened, which could be a favorable feature in the presence of dephasing.

\section{\label{sec:outlook}Conclusion and outlook}
The numerical approach discussed  here is capable of capturing  beyond mean-field effects and  entanglement build-up efficiently. For the  case of dipolar interactions presented here we have shown that the GDTWA gives an excellent approximation for short times and features correct quantum-thermalization in the long-time limit. This implies that in models which thermalize, e.g.~which are not many-body localized because of disorder, the method can give accurate predictions over all time-scales. The method can be applied to any model consisting of discrete coupled Hilbert spaces, such as spin systems featuring  $SU({\mathcal N})$ super-exchange interactions~\cite{gorshkovNatPhys2010}, the so called Subir-Ye-Kitaev models~\cite{subirSYK,SYK2016,anatoliSYK},  Bose-Hubbard models~\cite{Nagao_Semicla_2019}, etc. 

A general drawback of the TWA is the lack of a clear convergence parameter, and consequently the difficulty to assess the validity at long times. While this validity can be verified by comparison in analytical limits, with exact calculations for feasible system sizes, or by corroborating it with experimental data, it is desirable to have a more systematical validation procedure. With our GDTWA approach such a procedure could be achieved by systematically grouping multiple spin-$S$ particles (local Hilbert size of $\mathcal{N} = 2S+1$) into pairs of effectively larger spin-$\tilde{S}$ particles (local Hilbert space size $\mathcal{N}^2$) or larger clusters of $s_c$ spin-$S$ particles (local Hilbert space size $\mathcal{N}^{s_c}$). In such a way the exact quantum dynamics is preserved within the clusters, while only inter-cluster correlations are subjected to the TWA approximation. This comes at the cost of changing the number of classical equations from $N \times (\mathcal{N}^2-1)$ to $N_c \times (\mathcal{N}^{2s_c}-1)$, where $N_c$ is the number of clusters. By comparing the dynamics obtained for increasing cluster sizes $s_c$ (the simulation would hypothetically become exact in the extreme limit $N=s_c$) and different cluster choices, one could benchmark the convergence of the phase space method in regimes where no other methods are available for verification. For example, such an approach was recently suggested and analyzed in Ref.~\cite{anatolicluster2018}, however without utilizing initial discrete distributions.

\subsection*{Acknowledgements}
We thank Francesca Ferlaino's Er group and Bruno Laburthe-Tolra's Cr group and Chunlei Qu  for fruitful discussions. We also thank Asier Pineiro and Itamar Kimchi for reviewing the manuscript. {\bf Funding:} B.Z. is supported by the NSF through a grant to ITAMP. A.M.R is supported by the AFOSR grant FA9550-18-1-0319 and its MURI Initiative, by the DARPA and ARO grant W911NF-16-1-0576, the DARPA DRINQs grant, the ARO single investigator award W911NF-19-1-0210,  the NSF PHY1820885, NSF JILA-PFC PHY1734006 grants, and by NIST. J.S. is supported by the French National Research Agency (ANR) through the Programme d'Investissement d'Avenir under contract ANR-11-LABX-0058\_NIE within the Investissement d’Avenir program ANR-10-IDEX-0002-02, by IDEX project ``STEMQuS'', and through computing time at the Centre de calcul de l'Universit\'e de Strasbourg.

\appendix
\section{Initial states that can be represented within the GDTWA
}\label{app:diagonal_state_exact}

For the time-evolution in our GDTWA approach, correlations of GGMs such as $\llangle\lambda^i_\mu (0) \lambda_\nu^j (0)\rrangle$, $\llangle\lambda^i_\mu (0) \lambda_\nu^j(0) \lambda_\kappa^k (0)\rrangle$ are important. Here, we will show that for diagonal states the quantum mechanical correlations of the initial state are perfectly reproduced by our discrete distributions from Eq.~\eqref{eq:ddist}.  Correlations between sites trivially factorize since we assume a product state, and the lack of inter-site correlations is correctly re-produced by our discrete distributions~\eqref{eq:ddist}, which also factorize between sites. Here, we thus focus on a single spin (site indices are dropped for simplicity). By diagonal state $\ket{\alpha_0}$ with $\alpha_0 = 1,2,\dots,\mathcal{N}$, we mean any eigenstate of the diagonal GGMs defined in Eqs.~\eqref{eq:Ds}, which are eigenstates of $\hat S_z$.

We first note that for diagonal states, any power of a single GGM is fully re-produced by an average over our distribution since for $\hat \rho = \ket{\alpha_0}\bra{\alpha_0}$,
\begin{align}
{\rm Tr}\big[\hat \rho (\Lambda_\mu)^{n_{\mu}} \big] 
&= {\rm Tr}\Big[\ket{\alpha_0}\bra{\alpha_0} \Big(\sum_{a_\mu} a_\mu\ket{a_\mu} \bra{a_\mu}\Big)^{n_{\mu}}  \Big] \nonumber \\
&= \sum_{a_\mu} |\braket{a_\mu|\alpha_0}|^2(a_\mu)^{n_\mu} \nonumber \\
&= \sum_{a_\mu} p_\mu(\{\lambda_\mu = a_\mu\}) (\lambda_\mu)^{n_\mu}  
= \llangle (\lambda_\mu)^{n_\mu} \rrangle ,
\end{align} 
where we used the expansion of the GGM into the orthonormal eigenbasis $\Lambda_\mu = \sum_{a_\mu} a_\mu \ket{a_\mu} \bra{a_\mu}$. Since the discrete distributions defined in Eq.~\eqref{eq:ddist} are furthermore independent for different $\mu$, correlations between different GGMs always factorize:
\begin{align}
\llangle (\lambda_1)^{n_1} (\lambda_2)^{n_2} \dots (\lambda_\mathcal{D})^{n_{\mathcal{D}}} \rrangle=\llangle (\lambda_1)^{n_1} \rrangle \llangle (\lambda_2)^{n_2}  \rrangle\dots \llangle (\lambda_\mathcal{D})^{n_{\mathcal{D}}} \rrangle
\end{align} 
with arbitrary powers $n_1, n_2, \dots ,n_{\mathcal{D}}\geq 0$. We will show that this property also holds for correlations for the diagonal quantum state.

We proceed by showing that for $\mu \neq \nu$
\begin{align}
  \bra{\alpha_0}\mathcal{S} \Lambda_\mu  \Lambda_\nu \ket{\alpha_0}  &\equiv \frac{1}{2} \big( \bra{\alpha_0}   \Lambda_\mu \Lambda_\nu \ket{\alpha_0} + \bra{\alpha_0} \Lambda_\nu \Lambda_\mu  \ket{\alpha_0}\big) \nonumber\\
  &= \text{Re}  \bra{\alpha_0}\Lambda_\mu  \Lambda_\nu \ket{\alpha_0} 
    \nonumber\\
    &=\bra{\alpha_0}\Lambda_\mu \ket{\alpha_0} \bra{\alpha_0}\Lambda_\nu \ket{\alpha_0}\label{eq:fac2lam}
\end{align}
The symmetrization super-operator, $\mathcal{S}$, is necessary since generally different GGMs cannot be measured simultaneously as they do not always commute. Note that \eqref{eq:fac2lam} is always true if $\ket{\alpha_0}$ is an eigenstate of $\Lambda_{\mu}$ and $\Lambda_{\nu}$. This is trivially the case for all diagonal matrices, and therefore for all diagonal GGMs from Eqs.~\eqref{eq:Ds}. Moreover, note that for an arbitrary diagonal state $\ket{\alpha_0}$ there are only $2(\mathcal{N}-1)$ GGMs for which $\ket{\alpha_0}$ is not an eigenstate,
\begin{align}
	R^{\rm \alpha_0}_\xi &\equiv \frac{1}{\sqrt{2}} \big( \ket{\xi}\bra{\alpha_0} +  \ket{\alpha_0}\bra{\xi} \big) \quad \text{and} \quad \label{eq:al_eig_off}\nonumber\\
	I^{\rm \alpha_0}_\xi& \equiv \frac{1}{\sqrt{2}\mi} \big( \ket{\xi}\bra{\alpha_0} -  \ket{\alpha_0}\bra{\xi} \big).
\end{align}
with $\xi = 1,2,\dots,\alpha_0-1, \alpha_0+1,\dots,\mathcal{N}$. For all other off-diagonal GGMs, $\ket{\alpha_0}$ is an eigenstate with zero eigenvalue. We therefore have to only show that Eq.~\eqref{eq:fac2lam} also holds if at least one of the matrices  $\Lambda_{\mu}$ and $\Lambda_{\nu}$ is from Eqs.~\eqref{eq:al_eig_off}. It is straightforward to compute that for $\mu \neq \nu$
\begin{align}
   \bra{\alpha_0} \tilde \Lambda_\mu R^{\rm \alpha_0}_\xi  \ket{\alpha_0} &= 0 = \bra{\alpha_0} \tilde \Lambda_\mu \ket{\alpha_0}  \bra{\alpha_0}R^{\rm \alpha_0}_\xi  \ket{\alpha_0} \\
      \bra{\alpha_0} \tilde \Lambda_\mu I^{\rm \alpha_0}_\xi  \ket{\alpha_0} &= 0 = \bra{\alpha_0} \tilde \Lambda_\mu \ket{\alpha_0}  \bra{\alpha_0}I^{\rm \alpha_0}_\xi  \ket{\alpha_0} \\
   \bra{\alpha_0} R^{\rm \alpha_0}_\xi I^{\rm \alpha_0}_\eta  \ket{\alpha_0} &= 0 = \bra{\alpha_0} R^{\rm \alpha_0}_\xi  \ket{\alpha_0}  \bra{\alpha_0}I^{\rm \alpha_0}_\eta  \ket{\alpha_0} \nonumber\\&~~~~~~\quad\text{for}\quad \eta \neq \xi\\
    \text{Re}  \bra{\alpha_0} R^{\rm \alpha_0}_\xi I^{\rm \alpha_0}_\xi  \ket{\alpha_0} &= \text{Re}\left(\frac{1}{2\mi}\right) = 0 \nonumber\\&= \bra{\alpha_0} R^{\rm \alpha_0}_\xi  \ket{\alpha_0}  \bra{\alpha_0}I^{\rm \alpha_0}_\eta  \ket{\alpha_0}, 
\end{align}
where $\tilde \Lambda_\mu$ is any GGM not from \eqref{eq:al_eig_off}. Therefore, \eqref{eq:fac2lam} is correct. Additionally, we can use the fact that any power of a GGM from Eqs.~\eqref{eq:Rs}, Eqs.~\eqref{eq:Is}, and Eqs.~\eqref{eq:Ds} is either a diagonal matrix and trivially has $\ket{\alpha_0}$ as eigenstate, or it is proportional to itself, therefore also ($\mu\neq \nu$)
\begin{align}
    \bra{\alpha_0}\mathcal{S} (\Lambda_\mu)^{n_\mu}  (\Lambda_\nu)^{n_\nu} \ket{\alpha_0}  
    = \bra{\alpha_0} (\Lambda_\mu)^{n_\mu} \ket{\alpha_0} \bra{\alpha_0}(\Lambda_\nu)^{n_\nu} \ket{\alpha_0}\label{eq:fac2lampow}
\end{align}
holds for arbitrary integer powers of $n_\mu$ and $n_\nu$.

Lastly, it's straightforward to see that the pairwise property \eqref{eq:fac2lampow} also leads to a factorization of any other symmetrized product of multiple GGMs. To realize this, consider that
\begin{align}
	\bra{\alpha_0}(\Lambda_\mu)^{n_\mu} (\Lambda_\nu)^{n_\nu} (\Lambda_\kappa)^{n_\kappa} \ket{\alpha_0}
	\propto
	\bra{\alpha_0}\Lambda_\xi (\Lambda_\kappa)^{n_\kappa} \ket{\alpha_0}.
\end{align}
Here we used the group property of the GGMs, i.e.~that the product of any number of GGMs is again proportional to some other GGM, $\Lambda_\xi$. If $\Lambda_\xi$ commutes with $(\Lambda_\kappa)^{n_\kappa}$ we do not need to symmetrize, otherwise $\kappa \neq \xi$ and we can again use Eq.~\eqref{eq:fac2lampow} to see that $\bra{\alpha_0}\mathcal{S} \Lambda_\xi (\Lambda_\kappa)^{n_\kappa} \ket{\alpha_0}$ factorizes. From a repeated application of this argument it follows that
\begin{align}
   & \bra{\alpha_0}\mathcal{S} (\Lambda_1)^{n_1}  (\Lambda_2)^{n_2} \dots  (\Lambda_{\mathcal{D}})^{n_{\mathcal{D}}} \ket{\alpha_0} 
    \nonumber\\
    &=\bra{\alpha_0}(\Lambda_1)^{n_1} \ket{\alpha_0} \bra{\alpha_0}(\Lambda_2)^{n_2} \ket{\alpha_0} \bra{\alpha_0}(\Lambda_{\mathcal{D}})^{{n_{\mathcal{D}}}} \ket{\alpha_0} .\label{eq:diag_state_prop}
\end{align}


In particular, for states $\ket{\psi}$ that are not diagonal, but can be obtained from a diagonal state via any single-site unitary transformation ($\hat U^\dag \hat U = \id$), $\ket{\psi}=\hat U\ket{\alpha}$, we can transform the Hamiltonian into the rotated basis $\hat H'=\hat U^\dag\hat H \hat U$, and then obtain the equations of motion for GGMs in a similar way as Eq.~(\ref{eq:dLam}), with $\hat H$ replaced by $\hat H'$. The dynamics of an observable of interest $\hat O$ can the be found from evaluating $\hat O'=\hat U ^\dag\hat O  \hat U$. Since in this rotated basis the initial state becomes diagonal, the  expectation values of $\hat O'$ can be correctly described with the discrete sampling, and thus $\bra{\psi}\hat O\ket{\psi}$ in the original basis can be obtained. For the type i) initial states (used in the main text) with arbitrary $\theta$, we can use $\hat U=e^{i\theta\hat S^y}$ and perform sampling in the rotated basis with initial state $\ket{-S}$. For initial states that are a coherent superposition of several Zeeman levels, $\ket{\psi}=\sum_m c_m\ket{m}$, with $m=-S,-S+1,...S$, $c_m$ real numbers and $\sum_m c_m^2=1$, the unitary transformation can be constructed from $\hat U=\prod_{m=-S+1}^{S}\me^{-\mi{\rm arcsin}[c_m/\sqrt{1-\sum_{m'=S}^{m+1}c_{m'}^2}]\hat A_{-S,m}}$, with $\hat A_{-S,m}=\ket{-S}\bra{m}+\ket{m}\bra{-S}$ hermitian operators. The type ii) initial states considered in the main text constitute a special case in this category. In the case that initial states vary among particles, the overall transformation can be straightforwardly constructed accounting for spatial inhomogeneities: $\hat U_N=\otimes_j \hat U_j$ with $\hat U^\dag_j \hat U_j = \id_j$.

\section{Failure of state-tomography for Gaussian states}\label{app:gaussian_tom}

\begin{figure*}
\centering
\includegraphics[width=0.8\textwidth]{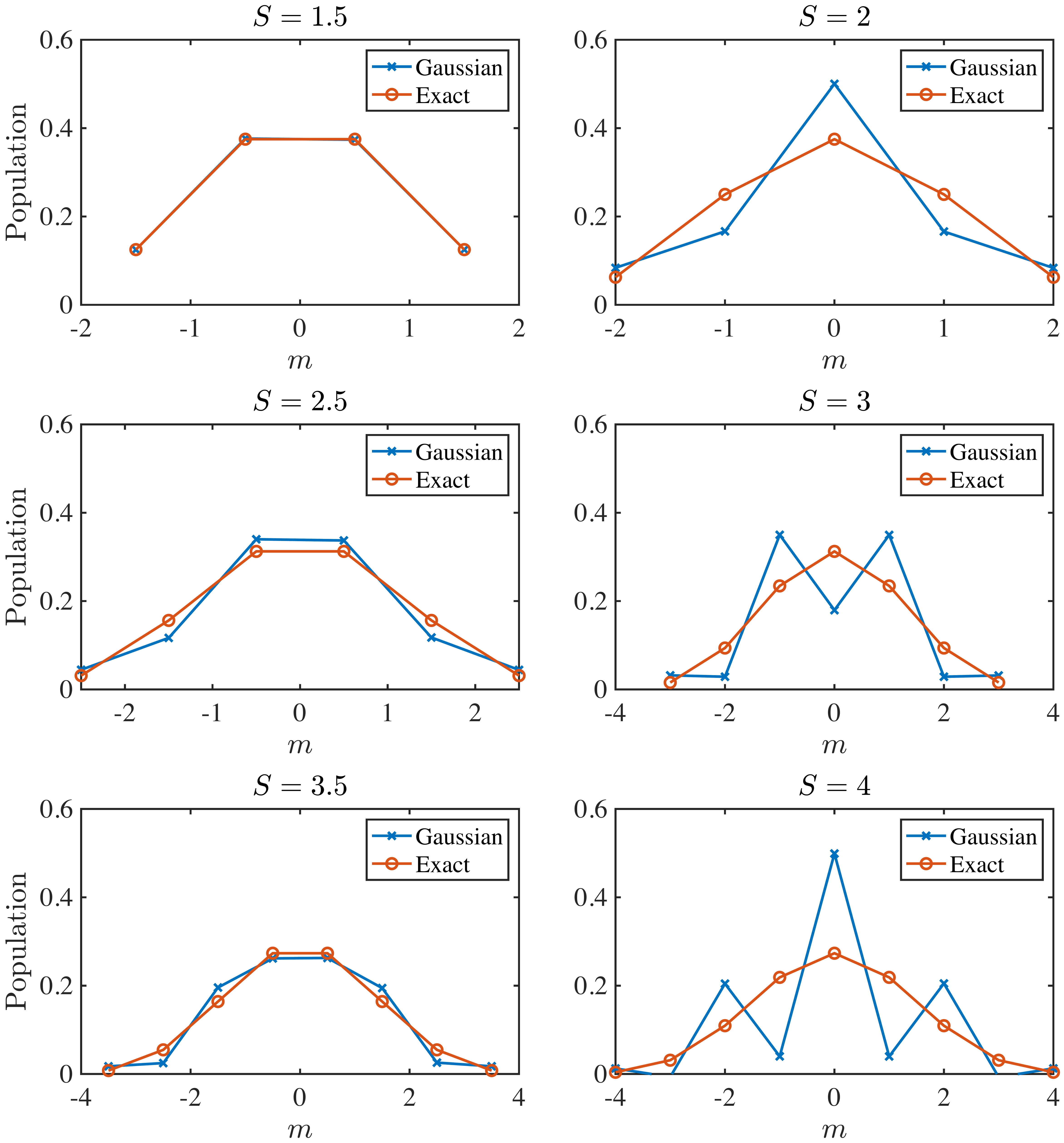}
\caption{\label{fig:gaussian_tom} The populations of Zeeman levels for an $x$ polarized spin coherent state, $|\braket{\rightarrow|m}|^2$, computed from both the exact quantum state and a Gaussian approximation to the spin coherent state. For $S\geq 2$ the Gaussian approximation clearly fails. 
}
\end{figure*}

Let's take a Gaussian state polarized along the $x$-axis, following from a $\pi/2$ rotation of the state from Eq.~\eqref{eq:gaussian_zstate} (site index dropped):
\begin{align}
W(S_x,S_y,S_z)=\frac{1}{\pi S}\me^{ -[S_y^2+S_z^2]/S}\delta(S_x-S),
\label{eq:gaussian_zstate_app}
\end{align}
This Gaussian distribution only reproduces the mean and the variance of the spin-operators correctly, but not the higher moments $\overline{S_z^n} = \int_{-\infty}^{\infty} d\bm{S}\, W(S_x,S_y,S_z) S_z^n,$ with $n>3$. For example, the first $4$ moments of the $z$ component are
$
\overline{S_z^0} = 1,
\overline{S_z^1} = 0,
\overline{S_z^2} = {S}/{2},
\overline{S_z^3} = 0,
\overline{S_z^4} = {3} S^2/4
$. In contrast, the exact quantum mechanical spin coherent state is
\begin{align}
\ket{\rightarrow} = \frac{1}{\sqrt{2^{2S}}} \sum_{i= 0}^{2S}  \sqrt{\binom{2S}{i}} \ket{i-S}
 \end{align}
 and gives rise to moments distributed according to a binomial distribution. In particular, the exact $4$-th moment is $\bra{\rightarrow} \hat S_z^4 \ket{\rightarrow} = {3}S^2/4 - {S}/{4}$ and differs from the Gaussian one by $S/4$.
 
 \begin{figure*}
\centering
\includegraphics[width=0.9\textwidth]{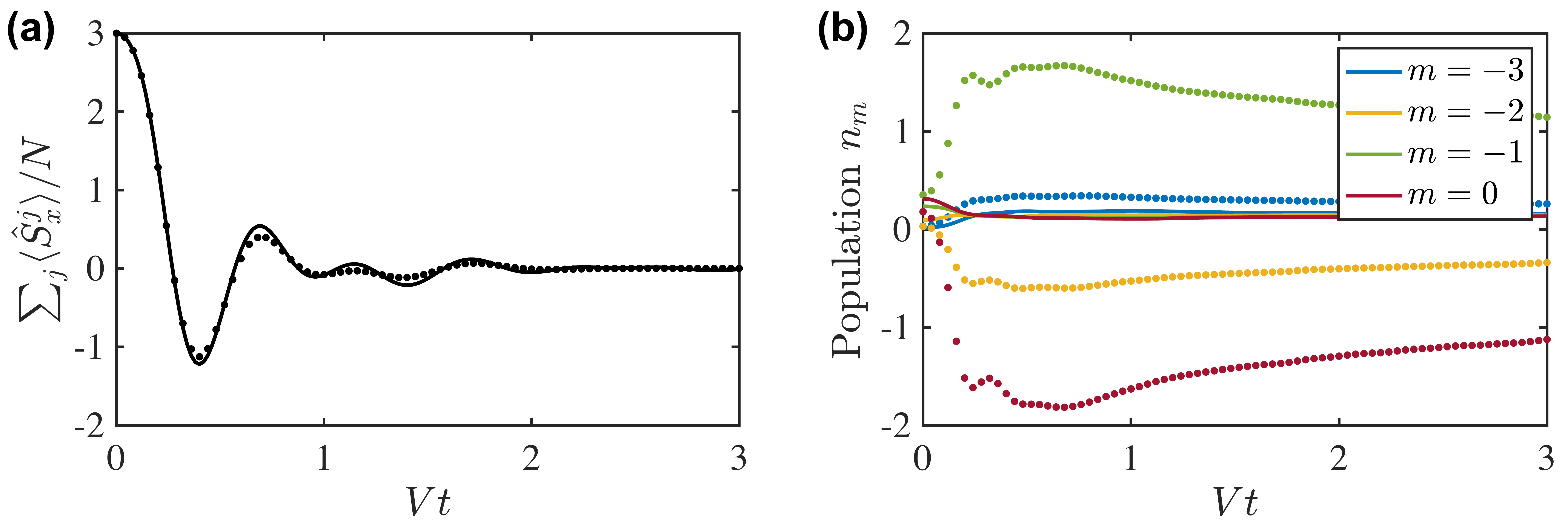}
\caption{\label{fig:ctwa_fail} Same simulation as in  Fig.~\ref{fig:ed}(a) ($S=3$). Here we compare a TWA simulation with a continuous Gaussian sampling for the three spin-variables (dots) with ED (lines). While the continuous Gaussian approach can capture the evolution of the spin-variables very well (a) (shown: $\sum_j \langle \hat S_j^x \rangle/N$), the evolution of Zeeman level populations (b) [obtained from state tomography via expectation values for $(\hat S_j^z)^2$, $(\hat S_j^z)^3$, etc.] leads to wrong, even unphysical results.
}
\end{figure*}

 The differences in the higher order moments has profound consequences when performing a state-tomography for large spin density matrices. For example, the diagonal density matrix elements (i.e.~the populations) can be expressed by powers of the spin-$z$ matrix. In the following let's take as an example the ``central'' Zeeman state $m_c=0$ ($S$ even) or $m_c = +1/2$ ($S$ odd). Then it is straightforward to see that
 \begin{align}
S = \frac{1}{2}: \quad & \ket{m_c}\bra{m_c} = \frac{1}{2} \hat S_z^0 +  \hat S_z  \label{eq:popex05}\\
S = 1: \quad & \ket{m_c}\bra{m_c} =  \hat S_z^0 -  \hat S_z^2\\
S = \frac{3}{2}: \quad & \ket{m_c}\bra{m_c} =  \frac{9}{16} \hat S_z^0 +  \frac{9}{8} \hat S_z^1 - \frac{1}{4} \hat S_z^2 - \frac{1}{2} \hat S_z^3\\
S = 2: \quad & \ket{m_c}\bra{m_c} =  \hat S_z^0 -  \frac{5}{4} \hat S_z^2 + \frac{1}{4} \hat S_z^4.\label{eq:popex2}
\end{align}
We see that for $S\geq 2$ the 4th order moment starts to play a role if we want to evaluate the expectation values for the $\hat S_z^n$ operators from a probability distribution. In particular, assuming that we choose a Gaussian distribution, then for $S=2$
\begin{align}
|\braket{m_c|\rightarrow}|^2 \approx \overline{S_z^0}  - \frac{5}{4} \overline{S_z^2} + \frac{1}{4}\overline{S_z^4}  = \frac{1}{2},
\end{align}
while the exact population of the state $\ket{m_c}$ is
\begin{align}
|\braket{m_c|\rightarrow}|^2 = \frac{3}{8}.
\end{align}
This difference is significant and it does not vanish with increasing spin. Fig.~\ref{fig:gaussian_tom} demonstrates the failure of the Gaussian sampling  for obtaining Zeeman level populations for various $S\geq2$. Especially, the large integer spin cases are very badly re-produced by a Gaussian distribution. In general, this shows that already the diagonal elements are not faithfully reproduced from a Gaussian sampling for an initial state, and the same arguments also hold for off-diagonal density-matrix terms, which in the case of $S\geq 2$ will require expectation values of $\hat S_x^{\geq 4}, \hat S_y^{\geq 4}$. Note that in contrast, our discrete distribution for diagonal states (including spin coherent states) is exact and provides correct predictions for moments of all orders (see~\ref{app:diagonal_state_exact}).

Consequently, also an attempt to simulate the time-evolution of Zeeman level populations using the state-tomography from expectation values of powers of $\hat S_j^z$ gives wrong results for $S\geq 2$ as demonstrated in Fig.~\ref{fig:ctwa_fail}. There we perform the same simulation as in Fig.~\ref{fig:ed}(a). The Gaussian sampling for the three spin-variables can reproduce results for dynamics of the spin-components very well (shown here: $\sum_j \langle \hat S_x^j \rangle/N$). In contrast, here the time-dependence of the populations has to be constructed from the time-dependent evolution of observables corresponding to the powers $(\hat S_j^z)^\eta$ via expansions such as~\eqref{eq:popex2}. While already the initial value for the powers with $\eta>3$ are badly reproduced (Fig.~\ref{fig:gaussian_tom}), the sub-sequent time-evolution then gives rise to even unphysical evolution of populations, reaching negative values or values larger than one already at short times. Note that the inital state of Fig.~\ref{fig:ed}(b) cannot be  described by a Gaussian distribution.

\begin{figure*}
\centering
\includegraphics[width=0.9\textwidth]{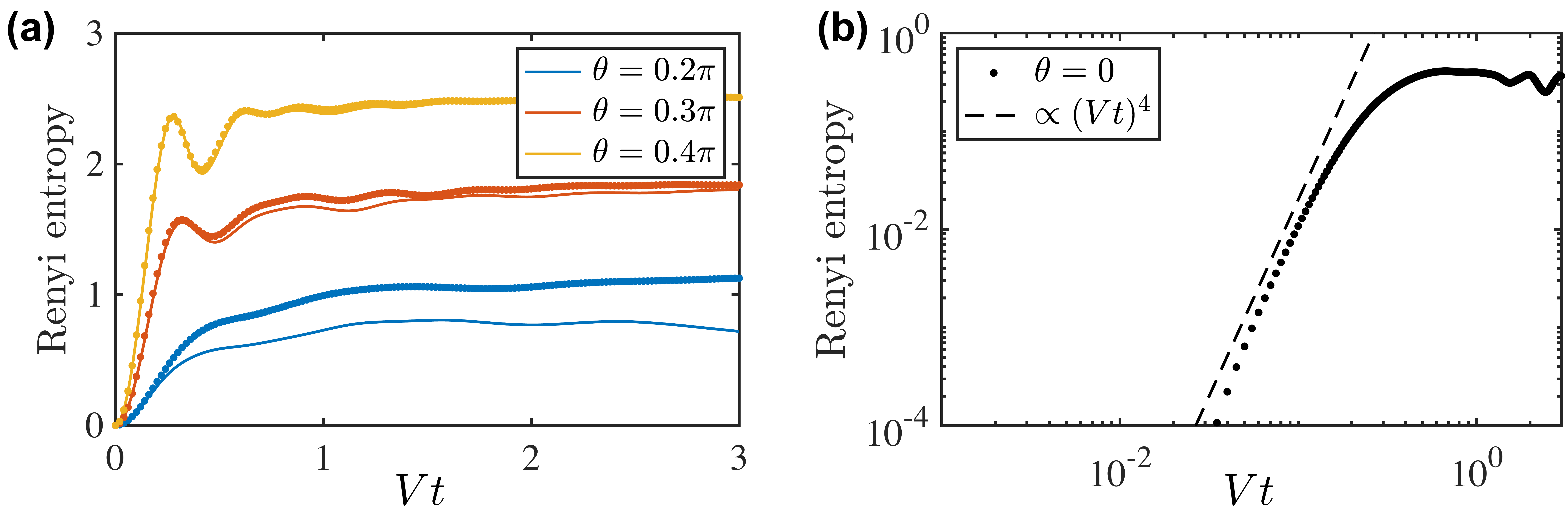}
\caption{\label{fig:edrenyitheta} (a) Second Renyi entropy obtained with GDTWA (dots) and ED (solid lines). The same parameters as in Fig.~\ref{fig:edrenyi}(a) are used. Here we consider tipping angles $\theta=0.2\pi$, $\theta=0.3\pi$ and $\theta=0.4\pi$. (b) Case of $\theta=0$ on a double-logarithmic scale. Here, ED shows always zero entropy.}
\end{figure*}

\section{Renyi entropy for smaller tipping angles ($\theta<\pi/2$)}\label{app:benchmarktheta}

In Fig.~\ref{fig:edrenyi}(a) we have shown that for $S=3$ and  tipping angle $\theta=\pi/2$ the GDTWA approach captures the second Renyi entropy nearly perfectly. Here to  further show  the validity of our approach, we provide results for several other tipping angles in Fig.~\ref{fig:edrenyitheta}. Except for $\theta=0.2\pi$, these  comparisons show a good agreement between the GDTWA and ED, suggesting that generally for $\theta$ not too small, the GDTWA approach works well for calculating the second Renyi entropy. Surprisingly, in the small $\theta$ case, in which even a pure mean-field simulations becomes valid to describe the short time dynamics, due to generally small build-up of entanglement, the GDTWA shows some intrinsic production of entanglement which is unphysical. We attribute this to some incorrect correlation build-up in higher-order correlations, which are incorrectly captured due to the TWA being a lowest order expansion (in $\hbar$). Indeed this is corroborated by Fig.~\ref{fig:edrenyitheta}(b), where we show the case of $\theta=0$. Quantum mechanically, one would not expect any dynamics in this case, since the initial state is an eigenstate of the Hamiltonian. However, the GDTWA still shows some dynamics. The deviation (difference with zero Renyi entropy) grows initially $\propto (Vt)^4$, which indicates that higher order corrections to the GDTWA are relevant. It is again worth pointing out, that nevertheless in the case of large entanglement build-up, these corrections seem to contribute only a relatively minor correction. This means that specifically in the regime where mean-field simulations fail, the GDTWA can provide very good estimations.

\section{Calculating spin squeezing in the GDTWA}\label{app:squeez}

\begin{figure*}
\centering
\includegraphics[width=0.55\textwidth]{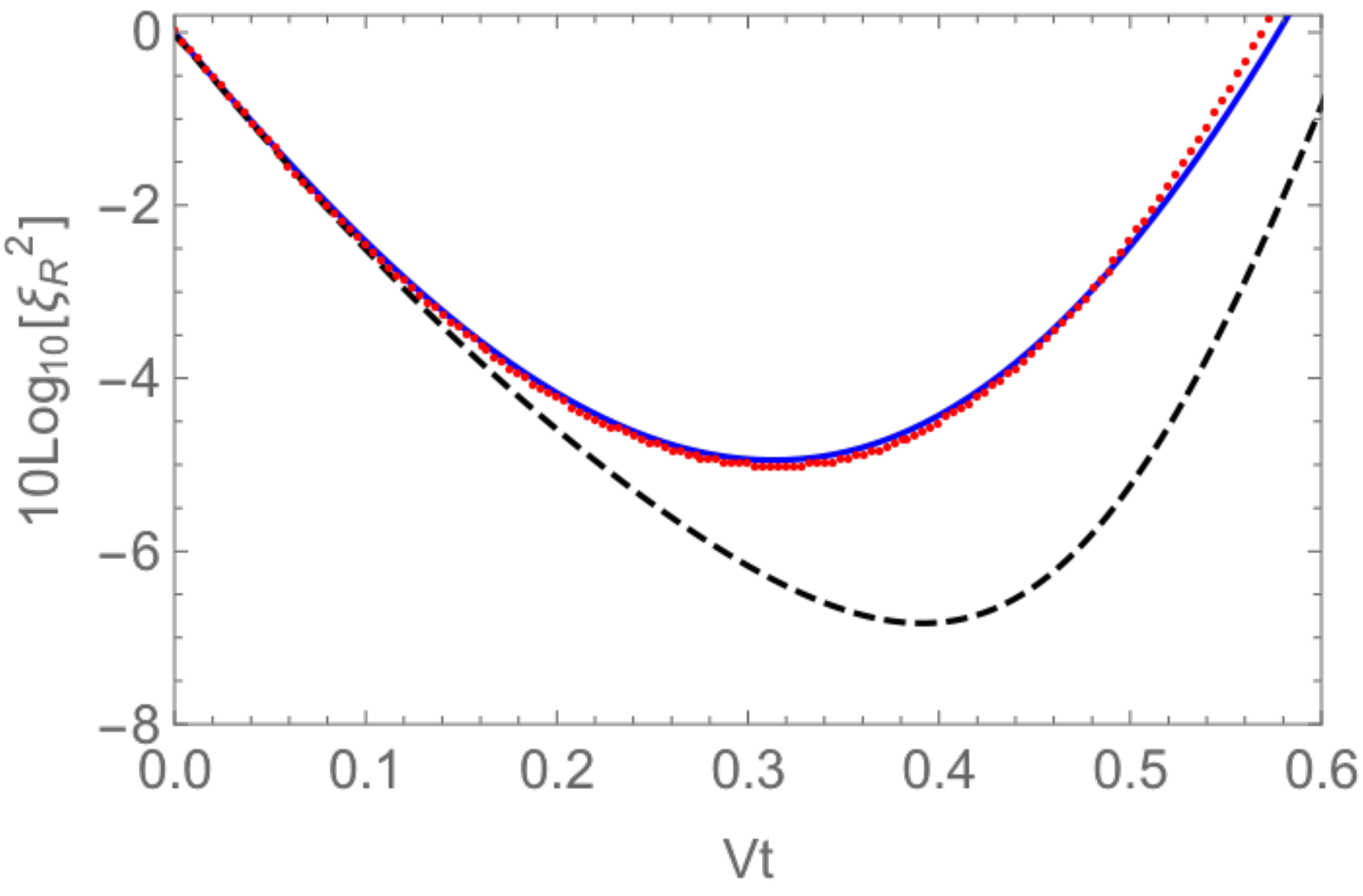}
\caption{\label{fig:onsite} Spin squeezing parameter calculated with  Eq.~\eqref{eq:gdtwas1} (red dots) and Eq.~\eqref{eq:gdtwas2} (black dashed line), and the comparion with ED (blue solid line). The parameters are the same as in Fig.~\ref{fig:squeezbmk}. }
\end{figure*}

To calculate the spin squeezing parameter $\xi_R^2$, we need to evaluate the expectation value of linear and quadratic operators such as $\hat S^x$ and $(\hat S^x)^2$. In the GDTWA, we first evaluate the expansion coefficients for the general operators in the GGM basis [Eq.~\eqref{eq:gen_op_exp}]:
\begin{align}
\hat S^x&=\sum_i\hat S^x_i=\sum_{i,\mu}c_{\mu}^i\Lambda^i_{\mu} \\
(\hat S^x)^2&= \sum_i(\hat S^i_x)^2+\sum_{i,j\neq i} \hat S^i_x\hat S^j_x \nonumber\\&= \sum_i \sum_\nu d_\nu^i \Lambda^i_{\mu} + \sum_{i,j\neq i} \sum_{\mu,\nu} c_\mu^i c_\nu^j \Lambda^i_{\mu}\Lambda^j_{\nu}. \label{eq:s2tot_exp}
\end{align}
Then, we approximate the quantum expectation values from averages over the classical trajectories: [Eq.~\eqref{eq:gdtwa}]:
\begin{align}
\langle \hat S^x \rangle &\approx \sum_{i,\mu}c_{\mu}^i\llangle\lambda^i_{\mu}(t)\rrangle \\
\langle (\hat S^x)^2 \rangle &\approx \sum_i \sum_\nu d_\nu^i \llangle\lambda^i_{\mu}(t)\rrangle + \sum_{i,j\neq i} \sum_{\mu,\nu} c_\mu^i c_\nu^j \llangle\lambda^i_{\mu}(t)  \lambda^j_{\nu}(t) \rrangle.
\label{eq:gdtwas2}
\end{align}

Alternatively, in the expansion of $(\hat S^x)^2$ in Eq.~\eqref{eq:s2tot_exp} one could also ignore the distinct treatment given to the diagonal terms, and instead compute the expectation value $\langle (\hat S^x)^2 \rangle$ as
\begin{align} 
  \langle (\hat S^x)^2 \rangle &\approx \sum_{i,j} \sum_{\mu,\nu} c_\mu^i c_\nu^j \llangle\lambda^i_{\mu}(t)  \lambda^j_{\nu}(t) \rrangle.
  \label{eq:gdtwas1}
\end{align}

In principle the first term in Eq.~\eqref{eq:gdtwas2} takes the single-spin part of the collective observable into account more precisely. Note, e.g.~that for $S=1/2$ it correctly reproduces the constant value for $ \langle \hat \sigma_x^2\rangle = \langle\id\rangle = 1$. Nevertheless, by comparisons with ED calculations, we find that in practice, especially for finite range interactions, Eq.~\eqref{eq:gdtwas1} describes the evolution of $\xi_R^2$ quantitatively better than Eq.~\eqref{eq:gdtwas2}. This is illustrated in Fig.~\ref{fig:onsite}, where a comparison with ED shows a nearly perfect agreement using Eq.~\eqref{eq:gdtwas1}, while a clear deviation on the logarithmic scale is visible when using Eq.~\eqref{eq:gdtwas2}. Therefore, in  the main text we adopted Eq.~\eqref{eq:gdtwas1} for evaluating spin squeezing.

\bibliographystyle{apsrev4-1}
\bibliography{gdtwa}

\end{document}